\pdfoutput=1
\documentclass[10pt,twocolumn,notitlepage]{article}

\usepackage{geometry}
\usepackage{multicol}
\usepackage{titling}
\usepackage{titlesec}
\titleformat*{\section}{\normalsize\bfseries}
\titleformat*{\subsection}{\normalsize \itshape}
\titleformat*{\subsubsection}{\normalsize \itshape}
 \usepackage[
backend=biber,
style=ieee,
]{biblatex}
\usepackage{balance}
\usepackage{amssymb}
\usepackage{amsmath,amsfonts}
\usepackage{algorithmic}
\usepackage{algorithm}
\usepackage{array}
\usepackage{textcomp}
\usepackage{stfloats}
\usepackage{url}
\usepackage{verbatim}
\usepackage{graphicx}
\usepackage[hidelinks]{hyperref}
\usepackage{subcaption}
\usepackage[affil-it]{authblk}
\usepackage{mathptmx}
\usepackage{amsthm}
\newtheorem{lemma}{Lemma}[]
\newtheorem{theorem}{Theorem}[]
\theoremstyle{definition}
\newtheorem{remark}{Remark}[]
\newtheorem{assumption}{Approximation}[]

\newtheorem{constraint}{Constraint}
\usepackage{setspace}
\usepackage{tikz}
\usepackage{flowchart}
\usetikzlibrary{arrows}
\usepackage{textcomp}
\usetikzlibrary{shapes,arrows,positioning,calc}
\tikzstyle{startstop} = [rectangle, rounded corners, minimum width=2cm, minimum height=1cm,text centered, draw=black]
\tikzstyle{process} = [rectangle, minimum width=2cm, minimum height=1cm, text centered, draw=black]
\tikzstyle{decision} = [diamond, minimum width=0.5cm, minimum height=0.5cm, text centered, draw=black]
\tikzstyle{arrow} = [thick,->,>=stealth]

\begin{document}

\title{Distributed decentralized receding horizon control for very large-scale networks with application to satellite mega-constellations$^\dagger$}

\author[1]{Leonardo Pedroso$^{*,}$}
\author[1]{Pedro Batista}
\affil[1]{Institute for Systems and Robotics, Instituto Superior T\'ecnico, Universidade de Lisboa, Portugal}
\predate{}\postdate{}\date{}
\renewcommand\Affilfont{\itshape\small}
\providecommand{\keywords}[1]{\medskip\textbf{\textit{Index terms---}} #1 {\par\noindent\rule{\linewidth}{0.5pt}}}
\renewenvironment{abstract}
{\rule{\linewidth}{0.5pt}\par\medskip
	{\bfseries\noindent\abstractname\par}\medskip}

\twocolumn[
\begin{@twocolumnfalse}
	\maketitle
	\begin{abstract}
	The implementation feasibility of control algorithms over very large-scale networks calls for hard constraints regarding communication, computational, and memory requirements. In this paper, the decentralized receding horizon control problem for very large-scale networks of dynamically decoupled systems with a common, possibly time-varying, control objective is addressed. Each system is assumed to be modeled by linear time-varying dynamics, which can be leveraged to approximate nonlinear systems about successive points of operation. A distributed and decentralized receding horizon control solution is put forward, which: i) takes communication delays into account; ii) allows local communication exclusively; and iii) whose computational and memory requirements in each computational unit do not scale with the dimension of the network. The scalability of the proposed solution enables emerging very large-scale applications of swarm robotics and networked control. This approach is applied to the orbit control problem of low Earth orbit mega-constellations,  featuring high-fidelity numerical simulations for the Starlink mega-constellation.
	\end{abstract}
\end{@twocolumnfalse}
]%
{
	\renewcommand{\thefootnote}%
	{\fnsymbol{footnote}}
	\footnotetext[1]{Corresponding author.\\ \textit{Email address}: \href{leonardo.pedroso@tecnico.ulisboa.pt}{leonardo.pedroso@tecnico.ulisboa.pt} (Leonardo Pedroso)\\
	$^\dagger$Published version (Available online on 4 Oct 2023). Cite as: L. Pedroso, P. Batista, Distributed decentralized receding horizon control for very large-scale networks with application to satellite mega-constellations, Control Engineering Practice, vol.~141, pp.~105728, 2023. \url{https://doi.org/10.1016/j.conengprac.2023.105728}.}
}
	
\section{Introduction}\label{sec:intro}
\subsection{Motivation}

The interest in decentralized control and estimation has been growing over the past decades since it enables technology that relies on large-scale networks of coupled systems. The well-known classical control solutions are centralized, which rely on computations on a single node and all-to-all communication between systems. The sheer communication and computational requirements of these solutions render its implementation infeasible as the dimension of the network increases. The emergence of applications in swarm robotics, such as unmanned aircraft formation flight \cite{Bereg15,thien2018}, unmanned underwater formations \cite{Viegas2012,yuan2017} and satellite formation control \cite{RussellCarpenter2002,ivanov2019}, as well as of large-scale networks of spatially scattered processes such as irrigation networks \cite{li2014offtake,prodan2017} and power distribution networks \cite{singh2016load,vlahakis2019distributed,sampathirao2021distributed} have accelerated the demand for efficient decentralized algorithms.

The decentralized control and estimation problems can be formulated as optimization problems subject to constraints that arise from the decentralized nature of the network. Despite the large research effort in this field, it remains an open problem even for linear time-invariant (LTI) systems due to its intractability \cite{blondel2000survey}. On top of that, a significant portion of the envisioned large-scale applications have underlying nonlinear dynamics, for which befitting solutions are even scarcer. Oftentimes, linearization techniques are employed, approximating the dynamics of the nonlinear system over successive operating points \cite{marinescu2010output}. This approach yields a more tractable linear time-varying (LTV) system for which decentralized control strategies have to be designed. On one hand, solutions for time-invariant dynamics can be synthesized offline. On the other hand, when it comes to the implementation of time-varying decentralized algorithms over very large-scale networks, more challenges are brought to light in addition to the intricacies of the decentralized problem. In fact, the time-varying nature of the problem calls for real-time synthesis, which requires additional communication, computational, and memory requirements. Furthermore, these resources are often very limited, which poses heavy constraints to the design problem. In fact, the solutions must cope with these limiting factors by distributing the load of the overall algorithm evenly among the systems that make up the network. For these reasons, a distributed and decentralized solution ought to be sought.

\subsection{Scope}
The focus of this paper is to design a decentralized and distributed receding horizon control (RHC) solution for very large-scale networks of dynamically decoupled systems with a common control objective. The problem is formulated considering decoupled LTV dynamics for each system and coupled LTV tracking output dynamics expressed in a generic, possibly time-varying, topology to model the  network-wise control objective. Severe design constraints are levied to ensure the feasibility of its real-time implementation on a very large-scale.



\subsection{Related Literature}
Note that, as aforementioned, a control solution for a network of systems with decoupled nonlinear dynamics and coupled nonlinear tracking outputs can be obtained from this formulation making use of linearization techniques. {For a comprehensive overview of the approaches to distributed RHC, see \cite{bemporad2010decentralized}.} \cite{farhood2015distributed} reduce the finite-horizon regulator problem of a network of interconnected LTV systems into a sequence of linear matrix inequalities (LMIs). However, even though the synthesized controller can be implemented in a decentralized framework with a topology inherited from the plant, the synthesis of the controller must be performed in real-time in a single node. Since the computational and memory requirements of this solution cannot be distributed across the systems in the network, the computational, memory, and communication burden render such solution unfeasible in practice for very large-scale networks. Recently, \cite{pedroso2021discretecontrol} address the problem of designing a decentralized control solution for a network of agents modeled by LTV dynamics. In this work, a well-performing decentralized solution leveraging convex relaxation is presented, which is validated in a network of nonlinear systems employing linearization techniques. Nevertheless, similarly to \cite{farhood2015distributed}, the real-time controller synthesis cannot be computed distributively. Thus, albeit these two works propose decentralized strategies, their real-time synthesis is centralized, whose communication, computational, and memory requirements become unbearable as the dimension of the network increases. This challenge has been noted and addressed to a great extent for the extended Kalman filter (EKF) problem over very large-scale networks of mobile robots. In this context, partially distributed solutions have been proposed relying on centralized-equivalent frameworks \cite{dai2016junction}, bookkeeping \cite{leung2009decentralized}, and covariance intersection and split covariance intersection methods \cite{carrillo2013decentralized, wanasinghe2014decentralized}. Nevertheless, for each computational unit associated with a system, the communication, memory, and computational requirements, respectively, of these solutions scale with the dimension of the network. Thus, these are not suited to the envisioned very large-scale applications. A promising step towards efficient distributed solutions has been made by \cite{luft2016recursive,luft2018recursive}. They propose a decentralized method that relies on an approximation of the covariance between the estimation error of each pair of systems, for a general network, which can be computed distributively and supports asynchronous communication and measurements. Although the computational and communication burden of every system does not grow with the dimension of the network, the memory requirements do. Recently, the authors proposed a distributed decentralized EKF \cite{pedroso2023distributed} whose communication, memory, and computational requirements of each computational unit do not grow with the dimension of the network. In this paper, an approach similar to \cite{pedroso2023distributed} is employed. Even though the control-estimation duality usually allows to establish analogous results, the intricacies of distributed and decentralized RHC do not allow for a straightforward extension. {Unlike the estimation problem, research into distributed RHC schemes for decoupled LTV systems is rather limited and focuses mainly on particular control problems. For instance, in \cite{bemporad2011decentralized} a solution is presented for the particular case of a formation of unmanned aerial vehicles. Nevertheless, some results for decoupled nonlinear systems have already matured. Although these are designed in a distributed scheme, because of the nonlinear dynamics they, generally, rely on local communication followed by the numerical solution of local optimization problems in real-time. A distributed RHC solution in continuous-time with stability guarantees is proposed in \cite{dunbar2004distributed}. A decentralized RHC scheme suitable for leader-follower topologies is presented in \cite{richards2004decentralized,richards2004decentralized2}. Another} very interesting distributed and decentralized approach to the RHC problem over networks of decoupled {nonlinear} systems is proposed in \cite{keviczky2006decentralized}. Therein, {unlike in this work, a priori knowledge of the overall system equilibrium is assumed. Their approach is to divide the global optimization problem in several smaller problems} that concern each system and its neighborhood. At each time instant, each system solves a local RHC problem to find optimal inputs for itself and the systems in its neighborhood. Then, in each system, the optimal input of the first instant of the finite window concerned with that system is used. Note that, in this framework, the optimal input that a system $i$ predicts for another system $j$ in its neighborhood is, generally, different from the optimal input that system $j$ computes for itself. Sufficient stability conditions are derived as a function of this mismatch between optimal solutions. However, this work does not provide a bound on the mismatch between solutions as a function of the network topology and dynamics. Moreover, in this framework, to compute the control input at each time instant, each system ought to receive the state of every neighbor and only then proceed with the solution of the local receding horizon optimization problem. Thus, it is challenging to implement it in practice without introducing significant delays.




\subsection{Proposed approach}
{In this paper, we employ a decentralized framework in which each system is associated with a computational unit that computes its own control input making use of local communication and local state feedback exclusively.} To live up to the challenges of an implementation over a very large-scale network, heavy design constraints regarding communication, computational, and memory requirements are imposed. First, neither the data transmissions between systems nor floating-point computations are considered to be instantaneous, thus the distributed and decentralized solution must account for and cope with the inevitable communication and computation delays. Second, the number of communication links that a system establishes with other systems shall not grow with the dimension of the network, i.e., the communication complexity of each system must be $\mathcal{O}(1)$. Third, the quantity of data stored in each computational unit shall not grow with the number of systems in the network. Fourth, the overall computational demand of the control solution must be distributed across all computational units and the computational complexity of algorithms implemented in each one must grow with $\mathcal{O}(1)$ with the dimension of the network. To devise such a solution, the propagation of the contribution to the global tracking cost of the correlation between the states of the systems is decoupled. This is achieved as the result of an approximation that is introduced, similarly to \cite{luft2016recursive,luft2018recursive}. A novel framework is proposed whereby each system $\mathcal{S}_i$ keeps an estimate of the contribution to the global tracking cost of the correlation between every pair of systems whose tracking output is coupled with the state of $\mathcal{S}_i$. These are updated in the computational unit associated with $\mathcal{S}_i$, neglecting the influence on the global tracking cost of the correlation between systems whose tracking outputs do not depend on a common system whose tracking output is coupled with the state of $\mathcal{S}_i$. Although it may seem puzzling at first sight, in what follows, this approximation is formally detailed, its origin and logic are explored, and its role on the decoupling of the contributions of each system to the global tracking cost is made clear. 

In contrast with the vast majority of the RHC solutions in the literature, the proposed approach abides by the aforementioned very large-scale feasibility constraints. Furthermore, contrarily to the innovative approach in \cite{keviczky2006decentralized} that follows such constraints, the proposed solution does not assume knowledge about the global equilibrium of the network. This is key, for instance, for the application considered in this work, whereby the global equilibrium is unknown.






\subsection{Motivation for application in LEO mega-constellations} 

To evaluate the effectiveness of this solution and show its scalability to very large-scale networks, it is applied to the orbit control problem of low Earth orbit (LEO) satellite mega-constellations.

Despite the failure of most large LEO constellation projects during the 1990s, the increasing technological advances and demand for broadband connectivity led to a reawakening of these projects over the past decade. LEO communication systems have plenty of  advantages over medium Earth orbit (MEO) and geosynchronous Earth orbit (GEO) systems \cite{gavish1997leo}. With the current technology, they are unarguably a solution to meet the soaring demand for reliable low-latency, high capacity, global broadband connectivity. It is possible to notice a paradigm shift from the use of constellations of a small number of highly complex satellites to the employment of a very large number of smaller and simpler satellites that cooperate in large-scale networks. In fact, the term mega-constellation has been coined to designate these very large-scale constellations. A number of these projects have been under development, such as Telesat Lightspeed, OneWeb, Starlink, and Project Kuiper, some of which are starting to be deployed. Not surprisingly, the greatest concern is their economical viability.

Nevertheless, the aforementioned paradigm change has not yet been accompanied by a paradigm change from an operations standpoint. In fact, as detailed by \cite{zhan2020challenges}, the tracking telemetry and command (TT\&C) system projected for these constellations does not differ from the TT\&C system architecture employed for a single satellite. As a result, the constellation monitoring systems are still constituted by a main mission control center (MCC) with several ground terminals scattered across the Earth. This framework is unsuitable for mega-constellations since it is very challenging to implement in practice and expensive to maintain. In fact, all-to-all communication is required between all the satellites via the MCC to transmit large quantities of data over long distances, which introduces delays and requires high power consumption. If, for several reasons, namely geo-political factors, it is not possible to establish a direct link with a ground station over some areas, it is necessary to retransmit TT\&C data with inter-satellite links (ISL) via a path of satellites to an available ground station. This continuous flow of data introduces an uneven communication pressure on these links, which calls for complex protocols and introduces delays. These are only but a few reasons to illustrate the inadequacy of this classical paradigm, which wastes much needed resources such as power and bandwidth. Given the paramount importance of cutting costs for the viability of mega-constellations, there has to be a paradigm change as far as TT\&C architecture is concerned so that it is efficient and cost effective \cite{zhan2020challenges}. The management of satellite mega-constellations could be carried out in a decentralized TT\&C architecture. Orbit determination and control, as well as other low-level constellation operations, could be carried out cooperatively relying on local communication between satellites assured by ISL. The shift towards a decentralized cooperative paradigm does not require all-to-all communication, the computational load is shared among the satellites, and data is transmitted over much shorter distances. {Such a paradigm shift has self-evident benefits in terms of increased efficiency and cost-effectiveness.} Albeit well-performing, the current decentralized solutions either require all-to-all communication, neglect inevitable communication and computation delays, employ extensive measurement bookkeeping, or are not computationally scalable, which is unsuitable to the control problem of LEO mega-constellations. On the contrary, the distributed and decentralized algorithm put forward in this paper follows heavy constraints to cope with limited communication, computational, and memory resources of space systems, which is critical for the onboard implementation in a mega-constellation. 

In this paper, it is applied to the decentralized orbit control problem, allowing for the cooperative maintenance of the constellation shape. A novel approach that relies on a set of relative orbital elements is devised aiming for efficiency and fuel saving. The source code of a MATLAB implementation of the algorithm and of the numerical simulations is available as an example in the DECENTER Toolbox at {\small \url{https://decenter2021.github.io/examples/DDRHCStarlink/}}.



\subsection{Statement of contributions}

The main contributions of this paper are threefold. First, severe constraints regarding communication, computational, and memory requirements for the implementation of a RHC solution in real-time in a very large-scale are rigorously defined. Second, a novel distributed decentralized RHC method that abides by the very large-scale feasibility constraints is devised. Third, the scalability of the proposed methods is illustrated on a high-fidelity simulation of a very large-scale pressing application, for which, to the best of the authors' knowledge, no other feasible solutions exist in the literature.

\subsection{Paper outline}
{The structure of the paper is as follows}. In Section~\ref{sec:ProbStatement}, the decentralized control problem is formulated {together with the} communication, computational, and memory constraints that the control solution {ought to follow} to  {enable a real-time implementation in very large-scale systems}. In Section~\ref{sec:DEKF}, the proposed distributed and decentralized RHC algorithm is derived. In Section~\ref{sec:app_sats}, the algorithm put forward in this paper is applied to the orbit control problem of LEO mega-constellations. {In Section~\ref{sec:concl}, the principal findings and conclusions drawn from this paper are presented.}

\subsection{Notation}
Throughout this paper, the identity, null, and ones matrices, all of proper dimensions, are denoted by $\mathbf{I}$, $\mathbf{0}$, and $\mathbf{1}$, respectively. Alternatively, $\mathbf{I}_n$, $\mathbf{0}_{n\times m}$, and $\mathbf{1}_{n\times m}$ are also used to represent the $n\times n$ identity matrix and the $n\times m$ null and ones matrices, respectively. The entry $(i,j)$ of a matrix $\mathbf{A}$ is denoted by $[\mathbf{A}]_{ij}$. The column-wise concatenation of vectors $\mathbf{x}_1,\ldots,\mathbf{x}_N$ is denoted by $\mathrm{col}(\mathbf{x}_1,\ldots,\mathbf{x}_N)$ and $\mathrm{diag}(\mathbf{A}_1,...,\mathbf{A}_N)$ {represents the block diagonal matrix whose diagonal blocks are matrices} $\mathbf{A}_1, ..., \mathbf{A}_N$. The Kronecker delta is denoted by $\boldsymbol{\delta}_{ij}$. Given a symmetric matrix $\mathbf{P}$, $\mathbf{P}\succ\mathbf{0}$  and $\mathbf{P}\succeq\mathbf{0}$ are used to point out that $\mathbf{P}$ is positive definite and positive semidefinite, respectively. The Kronecker product of two matrices $\mathbf{A}$ and $\mathbf{B}$ is denoted by $\mathbf{A} \otimes \mathbf{B}$. The cardinality of a set $\mathcal{A}$ is denoted by $|\mathcal{A}|$. The Cartesian product of two sets $\mathcal{A}$ and $\mathcal{B}$ is denoted by $\mathcal{A} \times \mathcal{B}$. The modulo operation is denoted by $a\: \mathrm{mod}\:b$, which returns the remainder of the integer division of  $a \in \mathbb{N}$ by $b \in \mathbb{N}$. The greatest integer less than or equal to $x\in \mathbb{R}$ is denoted by $\lfloor x \rfloor$.

\section{Problem statement} \label{sec:ProbStatement}

The problem statement is introduced in three {stages}. {In a first instance,} in Section~\ref{subsec:ModelNetwork}, the models of all of the systems of the network are defined {and} concatenated to {express} a global model for the {whole} network. Second, in Section~\ref{subsec:filter_def}, the local RHC controllers are {defined} and the control problem is formulated for the global controller. Third, in Section~\ref{subsec:com_req}, the communication, computational, and memory constraints are {established}. {It is worth noting that this problem is formulated and tackled for a generic network of systems with LTV dynamics and LTV tracking output couplings that reflect a network-wise control objective. This approach does not rely on any further assumptions concerning the individual dynamics of each system, the nature of the tracking output function, or the topology of the tracking couplings.}


\subsection{Model of the network}\label{subsec:ModelNetwork}
Consider a network of $N$ systems, $\mathcal{S}_i$ with $i = 1,\ldots,N$, each associated with {a} computational unit, $\mathcal{T}_i$. Each system is dynamically decoupled and modeled by LTV dynamics. Each system has also an LTV tracking output, which is coupled with a set of other systems. The tracking outputs can express a control objective that is common to all the systems (for an example, see Section~\ref{sec:app_sats}). {Thus, the topology of the network is portrayed by the tracking output couplings between systems, which can be time-varying. It may be studied as a directed graph} $\mathcal{G}:=(\mathcal{V},\mathcal{E})$, {where $\mathcal{V}$ is the set of vertices and $\mathcal{E}$ is the set of directed edges.} An edge $e$ {directed from vertex $j$ towards vertex $i$} is denoted by $e = (j,i)$. {The in-degree of a vertex $i$ is denoted by $\nu_i^-$, which is the number of edges directed towards it, and its in-neighborhood is denoted by $\mathcal{D}^-_i$, which is the set of indices of the vertices from which such edges originate.} {Likewise, the out-degree of a vertex $i$ is denoted by $\nu_i^+$, which is the number of edges directed from it, and its out-neighborhood is denoted by $\mathcal{D}^+_i$, which is the set of indices of the vertices towards which such edges are directed.} In this {setting}, each system is represented by a vertex, {i.e.}, system $\mathcal{S}_i$ is represented by vertex $i$, and, if the tracking output of $\mathcal{S}_i$ {is coupled with} the state of $\mathcal{S}_j$, {this interconnection is expressed by an edge $e = (j,i)$.} {It is crucial to emphasize that the direction of the edge holds significant importance. For example, if} the tracking output of $\mathcal{S}_i$ depends on the state of $\mathcal{S}_j$ {it does not imply the converse necessarily.} {Moreover, since} the local goal of each system $\mathcal{S}_i$ is to drive the tracking output to zero, {henceforth it considered without any loss of generality that $i\in \mathcal{D}^-_i$, i.e., the tracking output of system $\mathcal{S}_i$ depends on its state.} Thus, each vertex of the topology graph is assumed to have a self-loop.

The dynamics of system $\mathcal{S}_i$ are modeled by the discrete-time LTV system
\begin{equation}\label{eq:localDynamics}
	\begin{cases}
		\mathbf{x}_i(k+1) = \mathbf{A}_i(k)\mathbf{x}_i(k)+ \mathbf{B}_i(k)\mathbf{u}_i(k)\\
		\mathbf{z}_i(k) = \sum_{j\in \mathcal{D}^-_i}\mathbf{H}_{i,j}(k)\mathbf{x}_j(k),
	\end{cases}
\end{equation}
where $\mathbf{x}_i(k)\in\mathbb{R}^{n_i}$ is the state vector, $\mathbf{u}_i(k)\in \mathbb{R}^{m_i}$ is the input vector, and $\mathbf{z}_i(t)\in\mathbb{R}^{o_i}$ is the tracking output vector, all of system $\mathcal{S}_i$; matrices $\mathbf{A}_i(k)$, $\mathbf{B}_i(k)$, and $\mathbf{H}_{i,j}(k)$ with $j\in \mathcal{D}^-_i$ are time-varying matrices, known in $\mathcal{T}_i$, that model the dynamics of system $\mathcal{S}_i$ and its tracking output couplings with the other systems in its in-neighborhood. Note that linearization techniques can be employed to approximate the dynamics of a nonlinear system with a nonlinear tracking output as an LTV system of the form of \eqref{eq:localDynamics}.

The global dynamics of the network can, then, be modeled by the discrete-time LTV system
	\begin{equation}\label{eq:globalDynamics}
		\begin{cases}
			\mathbf{x}(k+1) = \mathbf{A}(k)\mathbf{x}(k)+ \mathbf{B}(k)\mathbf{u}(k)\\
			\mathbf{z}(k) = \mathbf{H}(k)\mathbf{x}(k),
		\end{cases}
	\end{equation}
	where $\mathbf{x}(k):= \mathrm{col}(\mathbf{x}_1(k),\ldots,\mathbf{x}_N(k)) \in \mathbb{R}^{n}$ is the global state vector, $\mathbf{u}(k):= \mathrm{col}(\mathbf{u}_1(k),\ldots,\mathbf{u}_N(k)) \in \mathbb{R}^{m}$ is the global input vector, and $\mathbf{z}(k):= \mathrm{col}(\mathbf{z}_1(k),\ldots,\mathbf{z}_N(k)) \in \mathbb{R}^{o}$ is the global tracking output vector;  $\mathbf{A}(k):=\mathrm{diag}(\mathbf{A}_1(k),\ldots,\mathbf{A}_N(k))$ and $\mathbf{B}(k):=\mathrm{diag}(\mathbf{B}_1(k),\ldots,\mathbf{B}_N(k))$ model the dynamics of the global network; and $\mathbf{H}(k)$ is a block matrix whose block of indices $(i,j)$ is $\mathbf{H}_{i,j}(k)$, if  $j\in \mathcal{D}^-_i$, and $\mathbf{0}_{o_i\times n_j}$ otherwise.
	
	Before proceeding with the problem statement, it is worth pointing out that the network-wise control objective of virtually all large-scale networks can be expressed by sparse tracking couplings. In particular, $\nu_i^-$, the number of tracking output couplings of a system $\mathcal{S}_i$, is bounded and it is independent of $N$. {The innovative solution put forward herein exploits the sparsity of these couplings to enable} a distributed and decentralized RHC algorithm {subject to severe} communication, computational, and memory {constraints}. 
	
	\subsection{Decentralized receding horizon {control} problem}\label{subsec:filter_def}

	The goal of the proposed RHC controller is to regulate the global tracking output making use of local linear state feedback. {On one hand, in a centralized setting the global state of the network is accessible to every system at the cost of all-to-all communication via a central node.} {On the other hand, in a decentralized setting, it is not the case}: each system $\mathcal{S}_i$ only has access to a {fraction} of the global state. {Herein}, a fully decentralized configuration is {addressed, i.e., at every discrete-time instant $k$, only the states of the systems in the in-neighborhood of  $\mathcal{S}_i$ are accessible} to $\mathcal{T}_i$. {Section~\ref{subsec:com_req} explores this  property in greater depth.} The control input of $\mathcal{S}_i$ is, thus, of the form 
	\begin{equation}\label{eq:localFilter}
		\mathbf{u}_i(k) = -\sum_{j\in \mathcal{D}^-_i}\mathbf{K}_{i,j}(k)\mathbf{x}_j(k),
	\end{equation}
	where $\mathbf{K}_{i,j}(k)$ for $j\in \mathcal{D}^-_i$ are the controller gains of $\mathcal{S}_i$.
	
	
	The {aim is to design optimal controller gains with respect to a performance metric}. {It is important to remark that the gains of each system cannot be independently designed due to the tracking couplings between systems.} {As a result,} the local controllers are concatenated to define a global controller, which is used to formulate a global RHC problem. The global control input is given by
			\begin{equation}\label{eq:globalFilter}
				\mathbf{u}(k) = -\mathbf{K}(k)\mathbf{x}(k),
			\end{equation}
			where $\mathbf{K}(k)$ is the global gain matrix. Note that the global control law \eqref{eq:globalFilter} is equivalent to the concatenation of the local control laws \eqref{eq:localFilter} if and only if $\mathbf{K}(k)$ follows the sparsity pattern of block matrix $\mathbf{E}_{\mathcal{D}}$, whose block of indices $(i,j)$ is given by
			\begin{equation*}
				{\mathbf{E}_{\mathcal{D}}}_{\;i,j} = 
				\begin{cases}
					\mathbf{1}_{m_i\times n_j},& j\in \mathcal{D}^-_i\\
					\mathbf{0}_{m_i\times n_j},& j\notin \mathcal{D}^-_i.
				\end{cases}
			\end{equation*} This sparsity condition is denoted as $\mathbf{K}(k)\in \mathrm{Sparse}(\mathbf{E}_{\mathcal{D}})$, with 
			\begin{equation*}\label{}
				\begin{split}
					\mathrm{Sparse}(&\mathbf{E})  = \left\{[\mathbf{K}]_{ij}\in\mathbb{R}^{m\times n}: \right. \\ &\quad \left.[\mathbf{E}]_{ij} = 0 \implies  [\mathbf{K}]_{ij}= 0;\: i= 1,...,m, \:j=1,...,n \right\}.
				\end{split}
			\end{equation*}
			If all-to-all communication were possible, then $\mathbf{E}_{\mathcal{D}} = \mathbf{1}$, {corresponding to a centralized configuration.}
			
			{It is now possible to state the} global RHC problem. The goal is to minimize an infinite-horizon performance cost 
			\begin{equation*}
				\begin{split}
					J_{\infty} &:= \sum_{i=1}^N {J_{i}}_{\infty}\\
					&= \sum_{i=1}^N \sum_{\tau = 0}^{\infty}\left(\mathbf{z}_i^T(\tau)\mathbf{Q}_i(\tau)\mathbf{z}_i(\tau) + \mathbf{u}_i^T(\tau)\mathbf{R}_i(\tau)\mathbf{u}_i(\tau)\right),
				\end{split}
			\end{equation*}
			where $\mathbf{Q}_i(\tau) \succeq 0$ and $\mathbf{R}_i(\tau) \succ 0$ are known time-varying matrices of appropriate dimensions that weigh the local tracking output and input of each system $\mathcal{S}_i$, respectively. The proposed method consists of an approximation to the solution of the infinite-horizon problem above, considering multiple finite-horizon problems with an associated cost of the form 
			\begin{equation}\label{eq:cost_finite_local}
				\begin{split}
					J(k) &:= \sum_{i=1}^{N} J_i(k) \\ 
					&= \sum_{i=1}^{N}  \Bigg(\mathbf{z}_i^T(k+H)\mathbf{Q}_i(k+H)\mathbf{z}_i(k+H) \\ 
					&+ \sum_{\tau = k }^{k+H-1}  \left(\mathbf{z}_i^T(\tau)\mathbf{Q}_i(\tau)\mathbf{z}_i(\tau) + \mathbf{u}_i^T(\tau)\mathbf{R}_i(\tau)\mathbf{u}_i(\tau)\right) \Bigg),
				\end{split}
			\end{equation} 
			where $H \in \mathbb{N}$ denotes the length of the finite window. The extension of this problem to an infinite-horizon is achieved by making use of a scheme similar to model predictive control (MPC). One considers a finite window $\{k, ... , k+H\}$, with $H$ large enough so that the gains computed within that window converge to those that would be obtained if an arbitrarily large window was used. To reduce the computational load, $d\in \mathbb{N}$ gains may be used, instead of just one, defining a new window and computing the gains associated with it every $d$ time steps. Although the higher $d$ is, the less the computational load is, if too large a value of $d$ is chosen, a degradation of performance and robustness may occur.
			
			To formulate the problem globally,  \eqref{eq:cost_finite_local} can be rewritten as
					\begin{equation*}
						\begin{split}
							J(k) = & \; \mathbf{z}^T(k+H)\mathbf{Q}(k+H)\mathbf{z}(k+H) \\
							&+\sum_{\tau = k }^{k+H-1}\left(\mathbf{z}^T(\tau)\mathbf{Q}(\tau)\mathbf{z}(\tau) + \mathbf{u}^T(\tau)\mathbf{R}(\tau)\mathbf{u}(\tau)\right),
						\end{split}
					\end{equation*}
					where $\mathbf{Q}(\tau) := \mathrm{diag}\left(\mathbf{Q}_1(\tau),\ldots,\mathbf{Q}_N(\tau)\right)$ and $\mathbf{R}(\tau) := \mathrm{diag}\left(\mathbf{R}_1(\tau),\ldots,\mathbf{R}_N(\tau)\right)$.
					{Remark that devising a decentralized controller for a network of systems, whose dynamics are governed by the LTV system \eqref{eq:localDynamics}, is analogous to devising a controller \eqref{eq:globalFilter} for the global system \eqref{eq:globalDynamics} whereby the gain follows a sparsity pattern. The goal is to optimally design a sequence of gains that satisfy the sparsity pattern corresponding to a fully decentralized configuration. It results, for a finite-horizon, in the following optimization problem}
							\begin{equation}\label{eq:OptimizationProblemInf}
								\begin{aligned}
									& \underset{\begin{subarray}{c}\mathbf{K}(\tau)\in \mathbb{R}^{m\times n} \\\tau \in \{k,\ldots,k+H-1\} \end{subarray}}{\text{minimize}}
									& & \!\!J(k)\\
									& \quad \;\text{subject to}
									& & \!\!\mathbf{K}(\tau) \in \mathrm{Sparse}(\mathbf{E}_{\mathcal{D}}), \\
									& & & \!\!\mathbf{x}(\tau\!+\!1)\! =\! \left(\mathbf{A}(\tau)\!-\!\mathbf{B}(\tau)\mathbf{K}(\tau)\right)\mathbf{x}(\tau),\\
									&&& \!\!\tau = k,\ldots, k+H-1.
								\end{aligned}\!\!\!\!
							\end{equation}
							{The optimization problem  \eqref{eq:OptimizationProblemInf} poses a significant challenge since it is nonconvex even for networks with LTI dynamics. A viable solution is to relax the optimization problem, rendering it convex and making it possible to employ well-established optimization techniques. Nevertheless,  the relaxed solution is merely an approximation to the original problem, despite being optimal for the relaxed version. Therefore, a cautious relaxation approach is essential to minimize the deviation between the two solutions. This approach is employed to devise the control solution proposed in this work.}

							
							
							\subsection{Feasibility constraints}\label{subsec:com_req}
							{Naturally, the control solution developed to solve \eqref{eq:OptimizationProblemInf} ought to be feasible to be implemented in real-time in a decentralized configuration. Specifically, the computation of each gain  $\mathbf{K}_{i,j}(k)$, $j\in \mathcal{D}^-_i$, in $\mathcal{T}_i$ ought to abide by severe constraints regarding communication, computational, and memory requirements. In what follows, these constraints are detailed rigorously.}
							
							
							
							
							{First, a crucial point to consider is the synchronization of the data transmissions. On one hand, a variable stored in $\mathcal{T}_j$ at time instant $k$ that is required to perform a computation in  $\mathcal{T}_i$ at time instant $k$ would have to be transmitted instantaneously. These are designated as hard real-time transmissions, which require complex synchronization procedures. On the other hand, in a soft real-time transmission, the receiving computational unit only makes use of the transmitted data since, at least, the discrete time-instant that follows the instant of the transmission.} Therefore, soft real-time transmission are robust to communication delays up to the discretization time-step. One can readily point out that the definition of the local control input \eqref{eq:localFilter} requires a hard real-time transmission. In fact, $\mathbf{x}_j(k)$, with $j\in \mathcal{D}^-_i$, has to be instantaneously transmitted to $\mathcal{T}_i$, because it is known to $\mathcal{T}_j$ only at time instant $k$ and it is required in $\mathcal{T}_i$ at time instant $k$. For this case in particular, various techniques can be used to allow for a feasible implementation, since the state of a system is a small data transfer and it can be easily predicted over small time intervals. {As a result, henceforth, the} communication requirements are focused on the gain computation, for which hard real-time transmissions are not allowed. 
							

							{Second, each system can only establish a communication link with a limited number of other systems in the network. While some solutions in the literature rely on all-to-all communication, they are not scalable for very large-scale networks. Thus, it is necessary to restrict the number of communication links established with each systems to grow with $\mathcal{O}(1)$ with the dimension of the network, $N$, ensuring that the communication complexity is $\mathcal{O}(1)$.  It is crucial to emphasize that the communication limitations are enforced at the protocol level, i.e., governing the exchange of data between systems, rather than limiting the physical communication links. Specifically, it is not permitted for $\mathcal{T}_i$ to access data from $\mathcal{T}_k$ via intermediary systems that could retransmit the information.}

							{Third, each computational unit has limited memory. Therefore, the quantity of data stored in each one must not scale with the dimension of the network, i.e., the data storage complexity of each computational unit must grow with $\mathcal{O}(1)$ with $N$.}
							
							{Fourth, the implementation of the local control solution requires computational resources in each computational unit, which are also limited. As such, the burden of the computation of the global control algorithm ought to be shared among all computational units, such that each performs computations related with its associated system exclusively. Hence, the computational complexity of each computational unit ought to grow with $\mathcal{O}$(1) with $N$.}
							
							%
							%
							
							
							{The control solution must satisfy the following constraints.}
							
							\begin{constraint} \label{constraint:real-time}
								Hard real-time transmissions are not allowed for the synthesis of controller gains.
							\end{constraint}
							
							\begin{constraint} \label{constraint:com}
								The communication complexity of each computational unit {must grow with $\mathcal{O}(1)$ with $N$.}
							\end{constraint}
							
							\begin{constraint} \label{constraint:storage}
								The data storage complexity of each computational unit {must grow with $\mathcal{O}(1)$ with $N$.}
							\end{constraint}
							
							\begin{constraint}  \label{constraint:computational}
								The computational complexity of each computational unit {must grow with $\mathcal{O}(1)$ with $N$.}
							\end{constraint}


							\section{Distributed and Decentralized RHC}\label{sec:DEKF}
							
							The {aim} is to {devise} a decentralized control solution that solves the optimization problem \eqref{eq:OptimizationProblemInf} subject to the communication, memory, and computational Constraints~\ref{constraint:real-time}--\ref{constraint:computational} {detailed} in Section~\ref{subsec:com_req}, which are critical for a feasible implementation to very large-scale systems. The derivation of the distributed and decentralized RHC solution {is carried out in three stages}: i)~convex relaxation of {\eqref{eq:OptimizationProblemInf}}; ii)~decoupling of the gain synthesis procedure; and iii)~scheduling of the computations. In Section~\ref{subsec:analysis_alg}, the communication, memory, and computational requirements of the proposed solution are thoroughly analyzed in light of the constraints that were levied. {First, this approach is devised} for a time-invariant network topology and, in Section~\ref{sec:tv_topology}, it is extended to a time-varying topology.
							
							\subsection{Convex relaxation}

							{A similar approach to the one used for the unconstrained RHC methodology is employed relying on the convex relaxation of \eqref{eq:OptimizationProblemInf}.} As pointed out in Section~\ref{sec:intro}, \eqref{eq:OptimizationProblemInf} has already been addressed in \cite{pedroso2021discretecontrol} without taking the communication, memory, and computational Constraints~\ref{constraint:real-time}--\ref{constraint:computational} put forward in Section~\ref{subsec:com_req} into account, which are not satisfied by the solution proposed therein. Nevertheless, some interesting results in \cite{pedroso2021discretecontrol} can be leveraged to devise a distributed solution. Consider the following preliminary result.

							\begin{lemma}\label{lema:OS_global}
								Let $\mathbf{l}_j$ denote a column vector whose entries are all set to zero except for the $j$-th one, which is set to $1$. The solutions to the necessary condition for a constrained minimum of the Hamiltonian of \eqref{eq:OptimizationProblemInf} follow
										\begin{equation}\label{apAeq:solOSEq}
											\begin{cases}
												\mathbf{l}_j^T\left[\left(\mathbf{S}(\tau)\mathbf{K}(\tau)-\mathbf{B}^T(\tau)\mathbf{P}(\tau+1)\mathbf{A}(\tau)\right)\right.\\
												\quad\quad\quad \quad\quad\left.\mathbf{x}(\tau)\mathbf{x}^T(\tau)\right]\mathbf{l}_i = 0 &\!\!\!\!\!\!\!\!\!\!\!\!\!\!\!\!,[\mathbf{E}_{\mathcal{D}}]_{ji} \neq 0 \\
												\mathbf{l}_j^T\mathbf{K}(\tau)\mathbf{l}_i = 0 &\!\!\!\!\!\!\!\!\!\!\!\!\!\!\!\!, [\mathbf{E}_{\mathcal{D}}]_{ji} = 0,
											\end{cases}
										\end{equation}
										for $\tau = k ,\ldots, k+H-1$, where $\mathbf{P}(\tau)$, $\tau = k, \ldots, k+H$, is a symmetric positive semidefinite matrix given by
										\begin{equation}\label{eq:PLQR}
											\begin{cases}
												\mathbf{P}(k+H) = \mathbf{H}^T(k+H)\mathbf{Q}(k+H)\mathbf{H}(k+H)\\
												\mathbf{P}(\tau)\!=\!\mathbf{H}(\tau)^T\mathbf{Q}(\tau)\mathbf{H}(\tau)+\mathbf{K}^T(\tau)\mathbf{R}(\tau)\mathbf{K}(\tau)\\ \quad \quad \quad  \!\!+ \left(\mathbf{A}(\tau)\!-\!\mathbf{B}(\tau)\mathbf{K}(\tau)\right)^T\!\mathbf{P}(\tau\!+\!1)\!\left(\mathbf{A}(\tau)\!-\!\mathbf{B}(\tau)\mathbf{K}(\tau)\right)\!,\!\!\!\!\!\!\!\!
											\end{cases}
										\end{equation}
										and
										\begin{equation}\label{eq:defS}
											\mathbf{S}(\tau) := \mathbf{B}^T(\tau)\mathbf{P}(\tau+1)\mathbf{B}(\tau)+\mathbf{R}(\tau).
										\end{equation}
										Furthermore, it follows that 
												\begin{equation}\label{eq:OSoptCost}
													\begin{split}
														\mathbf{x}^T(\tau)\mathbf{P}(\tau)\mathbf{x}(\tau) = &\; \mathbf{z}^T(k+H)\mathbf{Q}(k+H)\mathbf{z}(k+H) \\
														+& \sum_{s = \tau }^{k+H-1}\left(\mathbf{z}^T(s)\mathbf{Q}(s)\mathbf{z}(s) + \mathbf{u}^T(s)\mathbf{R}(s)\mathbf{u}(s)\right).
													\end{split}
												\end{equation}	
											\end{lemma}
											\begin{proof}
												See \cite[Appendix A]{pedroso2021discretecontrol} .   
											\end{proof}
											
											Lemma~\ref{lema:OS_global} puts emphasis on the nonconvexity of \eqref{eq:OptimizationProblemInf} since there are, in general, multiple solutions to the necessary condition of a constrained minimum. Herein, the convex relaxation approach is to approximate a single solution to \eqref{apAeq:solOSEq}, not necessarily optimal, which follows
											\begin{equation}\label{apAeq:solOSEq_rlx}
												\begin{cases}
													\mathbf{l}_j^T\left[\mathbf{S}(\tau)\mathbf{K}(\tau)-\mathbf{B}^T(\tau)\mathbf{P}(\tau\!+\!1)\mathbf{A}(\tau)\right]\mathbf{l}_i = 0 &,[\mathbf{E}_{\mathcal{D}}]_{ji} \neq 0 \\
													\mathbf{l}_j^T\mathbf{K}(\tau)\mathbf{l}_i = 0 &, [\mathbf{E}_{\mathcal{D}}]_{ji} = 0.
												\end{cases}
											\end{equation}
											This convex relaxation procedure, denoted as one-step relaxation, is analyzed with detail in \cite{pedroso2021discretecontrol}, where it is shown to be equivalent to three different relaxation hypothesis. 
											
											\subsection{Gain synthesis decoupling}\label{subsec:synth_dec}
											
											Define a block decomposition of $\mathbf{P}(\tau)$ and $\mathbf{S}(\tau)$, whose blocks of indices $(i,j)$ are denoted by $\mathbf{P}_{i,j}(\tau)\in \mathbb{R}^{n_i\times n_j}$ and $\mathbf{S}_{i,j}(\tau)\in \mathbb{R}^{m_i\times m_j}$, respectively. {Making use of this block decomposition, one can rewrite \eqref{eq:PLQR} in a decoupled manner for each of the blocks of $\mathbf{P}(\tau)$ as a function of the local dynamics and tracking output matrices as}
											\begin{equation*}
												\mathbf{P}_{p,q}(k\!+\!H) = \!\!\!\!\sum\limits_{r \in \mathcal{D}^+_p \cap  \mathcal{D}^+_q} \!\!\!\!\mathbf{H}_{r,i}^T(k\!+\!H)\mathbf{Q}_{r}(k\!+\!H)\mathbf{H}_{r,j}(k\!+\!H)
											\end{equation*}\vspace{-0.5cm}
											and
											\begin{equation}\label{eq:cases_decoupled_P}
												\begin{split}
													\mathbf{P}_{p,q}&(\tau) = \sum_{r \in \mathcal{D}^+_p \cap  \mathcal{D}^+_q}^{\phantom{a}}\mathbf{H}_{r,i}^T(\tau)\mathbf{Q}_{r}(\tau)\mathbf{H}_{r,j}(\tau) \\
													&+\sum_{r \in \mathcal{D}^+_p \cap  \mathcal{D}^+_q}^{\phantom{a}}\mathbf{K}_{r,i}^T(\tau)\mathbf{R}_{r}(\tau)\mathbf{K}_{r,j}(\tau)\\
													&+\sum_{r\in \mathcal{D}^+_p}\sum_{s\in \mathcal{D}^+_q}\left(\mathbf{A}_p(\tau)\boldsymbol{\delta}_{pr}-\mathbf{B}_r(\tau)\mathbf{K}_{r,p}(\tau)\right)^T\\
													&\quad \quad \mathbf{P}_{r,s}(\tau+1)\left(\mathbf{A}_q(\tau)\boldsymbol{\delta}_{qs}-\mathbf{B}_s(\tau)\mathbf{K}_{s,q}(\tau)\right).
												\end{split}
											\end{equation}
											{One can also express the blocks of $\mathbf{S}(\tau)$ as a function of the blocks of $\mathbf{P}(\tau+1)$ as}
											\begin{equation*}
												\mathbf{S}_{i,j}(\tau) = \mathbf{B}_i^T(\tau)\mathbf{P}_{i,j}(\tau+1)\mathbf{B}_j(\tau) + \boldsymbol{\delta}_{ij}\mathbf{R}_j(\tau),
											\end{equation*}
											{which follows immediately from \eqref{eq:defS}.}
											{Moreover, leveraging the aforementioned block decomposition, the relaxed conditions \eqref{apAeq:solOSEq_rlx} of the feedback gains of the form $\mathbf{K}_{j,i}(\tau)$ can also be written in a decoupled manner, for each $i \in \{1,\ldots, N\}$, as} 
											\vspace{-0.2cm}\begin{equation}\label{eq:cases_decoupled_K}
												\!\!\!\!\begin{cases}
													\sum \limits_{p\in \mathcal{D}^+_i} \!\!\mathbf{S}_{j,p}(\tau)\mathbf{K}_{p,i}(\tau) -  \mathbf{B}_j^T(\tau)\mathbf{P}_{j,i}(\tau+1)\mathbf{A}_i(\tau) = \mathbf{0}, & j\in \mathcal{D}^+_i\\
													\mathbf{K}_{j,i}(\tau) = \mathbf{0}, & j\notin \mathcal{D}^+_i.\!\!\!
												\end{cases} 
											\end{equation}
											%
											For each set $\mathcal{D}_i^+$, let $\mathcal{D}_i^+ = \{p^i_1,\ldots,p^i_{|\mathcal{D}_i^+|}\}$. Then, concatenating {the feedback gains of the form $\mathbf{K}_{j,i}(\tau)$, with $j\in \mathcal{D}^+_i$, and combining the corresponding decoupled relaxed conditions of the first member of \eqref{eq:cases_decoupled_K}}, it follows that
											\begin{equation}\label{eq:OSgain}
												\tilde{\mathbf{K}}_i(\tau) = \tilde{\mathbf{S}}_i(\tau)^{-1}\tilde{\mathbf{P}}_i(\tau+1),
											\end{equation}
											where\vspace{-0.2cm}
											\begin{equation*}
												\tilde{\mathbf{K}}_i(\tau) := \begin{bmatrix}
													\mathbf{K}_{p^i_1,i}\\
													\vdots\\
													\mathbf{K}_{p^i_{|\mathcal{D}_i^+|},i}
												\end{bmatrix},
											\end{equation*} \vspace{-0.2cm}
											\begin{equation}\label{eq:def_Stilda}
												\tilde{\mathbf{S}}_i(\tau) := \begin{bmatrix}
													\mathbf{S}_{p^i_1,p^i_1} & \ldots & \mathbf{S}_{p^i_1,p^i_{|\mathcal{D}_i^+|}} \\
													\vdots & \ddots & \vdots \\
													\mathbf{S}_{p^i_{|\mathcal{D}_i^+|},p^i_1} & \ldots & \mathbf{S}_{p^i_{|\mathcal{D}_i^+|},p^i_{|\mathcal{D}_i^+|}}
												\end{bmatrix},
											\end{equation}
													and
													\begin{equation}\label{eq:def_Ptilda}
														\tilde{\mathbf{P}}_i(\tau\!+\!1) := \begin{bmatrix}
															\mathbf{B}_{p^i_1}^T(\tau)\mathbf{P}_{p^i_1,i}(\tau+1)\mathbf{A}_i(\tau)\\
															\vdots\\
															\mathbf{B}_{p^i_{|\mathcal{D}_i^+|}}^T(\tau)\mathbf{P}_{p^i_{|\mathcal{D}_i^+|},i}(\tau+1)\mathbf{A}_i(\tau)
														\end{bmatrix}.\;
													\end{equation}
													
													Note that, although \eqref{eq:OSgain} provides an expression to compute the gains of systems $\mathcal{S}_j$ with $j\in \mathcal{D}_i^+$ in relation to the state of system $\mathcal{S}_i$, it cannot be computed without full knowledge of $\mathbf{P}(\tau+1)$. This is due to the fact that the propagation of $\mathbf{P}_{j,i}(\tau)$ with $j\in \mathcal{D}^+_i$ in \eqref{eq:cases_decoupled_P} depends on several $\mathbf{P}_{r,s}(\tau+1)$ with $r\in \mathcal{D}^+_j$ and $s\in \mathcal{D}^+_i$, which in turn depends on blocks of $\mathbf{P}(\tau+2)$ that are even less coupled with $\mathcal{S}_i$, and so on. {Hence, this result is unfit to be implemented in a distributed framework. In what follows, under a thoughtful approximation, the computation of these gains is decoupled. Then, an algorithm is proposed to enable their distributed computation across the computational units of each system such that the communication, memory, and computational constraints detailed in Section~\ref{subsec:com_req} are followed.}
													
													\begin{assumption}\label{ass:approxP}
														Consider $\mathbf{P}_{p,q}(\tau)$, with $p \in \mathcal{D}^+_i$ and $q \in \mathcal{D}^+_i$ for some $i$, and $\mathbf{P}_{r,s}(\tau+1)$, with $r \in \mathcal{D}^+_p$ and $s \in \mathcal{D}^+_q$. In the decentralized algorithm put forward in this paper, $\mathbf{P}_{r,s}(\tau+1)$ is considered to be null in the computation of $\mathbf{P}_{p,q}(\tau)$ in the computational unit $\mathcal{T}_i$ if $(r,s)\notin \psi_i$, where
														\begin{equation}\label{eq:def_psi}
															\psi_i = \bigcup_{j\in \mathcal{D}^+_i}\phi_j,
														\end{equation}
														with
														\begin{equation}\label{eq:def_phi}
															\phi_i := \mathcal{D}^+_i  \times \mathcal{D}^+_i = \{(p,q)\in \mathbb{N}^2 : p\in \mathcal{D}^+_i  \land q \in \mathcal{D}^+_i\}.
														\end{equation}
													\end{assumption}
													
\begin{figure}[]
														\includegraphics[width=0.75\linewidth]{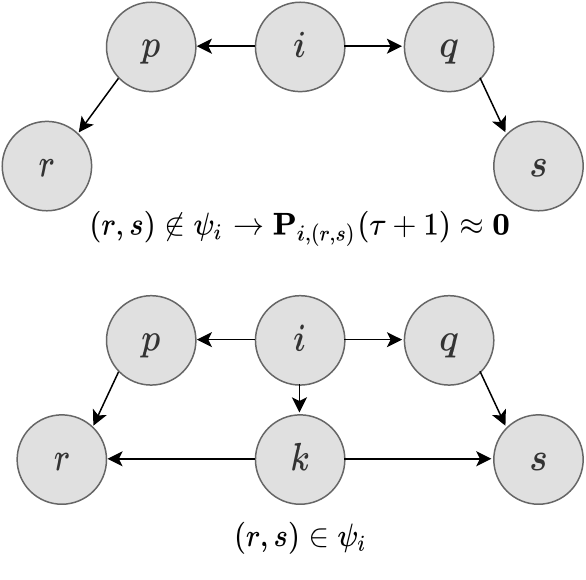}
														\centering
														\caption{Graphic illustration of Approximation~\ref{ass:approxP}.}
														\label{fig:assumption}
													\end{figure}

													The main result of this paper {follows from} Approximation~\ref{ass:approxP}. Next, it is argued that this approximation makes sense in the context of RHC of a large-scale network. Note that, from \eqref{eq:OSoptCost} in Lemma~\ref{lema:OS_global}, $\mathbf{P}_{r,s}(\tau)$ is a measure of the contribution of the correlation between the states of systems $\mathcal{S}_r$ and $\mathcal{S}_s$ to the global cost. {Figure~\ref{fig:assumption} depicts} the topology of Approximation~\ref{ass:approxP} in a graph. Intuitively, it is expected that the influence of $\mathbf{P}_{r,s}(\tau+1)$ is more dominant in the computation of $\mathbf{K}_{p,i}(k)$ for $p \in \mathcal{D}_i^+$ if both the states of $\mathcal{S}_r$ and $\mathcal{S}_s$ are coupled with the output of a system $\mathcal{S}_k$ that is coupled with the output of $\mathcal{S}_i$. {As a result, to enable a decoupled computation of the synthesis of each local controller,} each computational unit $\mathcal{T}_i$ keeps in memory and updates each $\mathbf{P}_{p,q}(\tau)$ with $(p,q)\in \phi_i$. Henceforth, the approximation of matrix $\mathbf{P}_{p,q}(\tau)$ that is stored and updated in $\mathcal{T}_i$ is denoted by $\mathbf{P}_{i,(p,q)}(\tau)$. Thus, making use of Approximation~\ref{ass:approxP}, \eqref{eq:cases_decoupled_P} becomes
													\begin{equation}\label{eq:cov_prop_ass}
														\begin{split}
															&\mathbf{P}_{i,(p,q)}(\tau) =\sum_{r \in \mathcal{D}^+_p \cap  \mathcal{D}^+_q}^{\phantom{a}}\mathbf{H}_{r,i}^T(\tau)\mathbf{Q}_{r}(\tau)\mathbf{H}_{r,j}(\tau) \\
															& \quad + \sum_{r \in \mathcal{D}^+_p \cap  \mathcal{D}^+_q}^{\phantom{a}}\mathbf{K}_{r,i}^T(\tau)\mathbf{R}_{r}(\tau)\mathbf{K}_{r,j}(\tau) \\
															&\quad + \sum_{r\in \mathcal{D}^+_p}\sum_{\substack{s \in \mathcal{D}^+_q\\(r,s)\in\psi_i}}\left(\mathbf{A}_p(\tau)\boldsymbol{\delta}_{pr}-\mathbf{B}_r(\tau)\mathbf{K}_{r,p}(\tau)\right)^T\\
															&\quad \quad \mathbf{P}_{\mathcal{D}^+_i,(r,s)}(\tau+1)\left(\mathbf{A}_q(\tau)\boldsymbol{\delta}_{qs}-\mathbf{B}_s(\tau)\mathbf{K}_{s,q}(\tau)\right),
														\end{split}
													\end{equation}
													where the subscript $\mathcal{D}_i^+$ in $\mathbf{P}_{\mathcal{D}^+_i,(r,s)}(\tau+1)$ indicates, by abuse of notation, that $\mathbf{P}_{\mathcal{D}^+_i,(r,s)}(\tau+1)$ is computed in $\mathcal{T}_k$ with $k\in \mathcal{D}_i^+$.
													Note that, with Approximation~\ref{ass:approxP}, the propagation of $\mathbf{P}(\tau)$ in \eqref{eq:cov_prop_ass} can be computed in a distributed manner. {Note} that $\mathbf{P}_{\mathcal{D}_i^+,(r,s)}(\tau+1)$, inside the summation in \eqref{eq:cov_prop_ass}, is not necessarily computed in $\mathcal{T}_i$, since only $\mathbf{P}_{i,(p,q)}(k|k)$, with $(p,q)\in \phi_i$, are updated in $\mathcal{T}_i$. Therefore, $\mathcal{S}_i$ has to receive $\mathbf{P}_{k,(r,s)}(\tau+1)$ through communication from a system $\mathcal{S}_k$, with $k \in \mathcal{D}_i^+$. 
													

													{Similarly to the analogous approximation for the estimation problem, it can be shown analogously that the approximation induced by Approximation~\ref{ass:approxP} is exact for certain topologies such as string, tree, and ring configurations \cite{pedroso2023distributed}[Lemma 2]. In such topologies, it is possible to distribute the computations of the globally synthesized controller among the computational units without disregarding any components of $\mathbf{P}(\tau+1)$. The proposed RHC algorithm, which follows from Approximation~\ref{ass:approxP}, is presented in the following result.}
													%
													
													
													
													\begin{theorem}[Distributed and decentralized RHC]\label{th:OS}
														The solution of the optimization problem \eqref{eq:OptimizationProblemInf} under the one-step convex relaxation \eqref{apAeq:solOSEq_rlx}, under Approximation~\ref{ass:approxP}, and subject to the communication, computational, and memory constraints detailed in Section~\ref{subsec:com_req} is given, for a time-invariant coupling topology, by the local gain computations presented in Algorithm~\ref{alg:OSDEKF}.
														\begin{proof}
															{The decoupled computation of each local gain is portrayed by \eqref{eq:OSgain}. Taking into account Approximation~\ref{ass:approxP}, the distributed computation of $\mathbf{P}(\tau)$ is carried out according to \eqref{eq:cov_prop_ass}. All variables that are required for these computations in each computational unit $\mathcal{T}_i$ are available through local communication with computational units in its out-neighborhood, i.e., $\mathcal{T}_j$ for $j\in \mathcal{D}_i^+$. Under the scheduling of the computations presented in Section~\ref{sec:scheduling_ti}, it is shown in Section~\ref{subsec:analysis_alg} that computational, and memory requirements of this algorithm satisfy the very large-scale feasibility constraints.}
														\end{proof}
													\end{theorem}

													\begin{algorithm*}[ht!]
														\caption{Distributed and decentralized RHC algorithm for the local gain synthesis of a new window of gains at time instant $k$ in computational unit $\mathcal{T}_i$, for a time-invariant coupling topology.}\label{alg:OSDEKF}
														\begin{algorithmic}
															\STATE
															\STATE{\textbf{Output:}} {$\mathbf{K}_{i,p}(\tau), \forall p\in \mathcal{D}_i^-, \tau = k,\dots,k+d-1$}
															\STATE \textbf{Step 1}: \textbf{Predict}: $\mathbf{A}_i(\tau), \mathbf{B}_i(\tau)$, $\mathbf{R}_i(\tau), \tau = k,\ldots,k+H-1$;
															\STATE \hspace{2.3cm} $\mathbf{H}_{i,p}(\tau), \forall p \in \mathcal{D}_i^-, \tau = k+1,\ldots, k+H$;\\
															\hspace{2.3cm} $\mathbf{Q}_{i}(\tau), \tau = k+1,\ldots,k+H$.
															\STATE \textbf{Step 2}: \textbf{Transmit}: $\mathbf{Q}_{i}(k+H)^{1/2}\mathbf{H}_{i,p}(k+H), \forall p\in \mathcal{D}_i^-$ to $\forall p\in \mathcal{D}_i^- \setminus \{i\}$.\\
															\textbf{Step 3}: \textbf{Receive}: $\mathbf{Q}_{p}(k+H)^{1/2}\mathbf{H}_{p,i}(k+H)$ from $\forall p \in \mathcal{D}^+_i \setminus \{i\}$.\\
															\textbf{Step 4}: \textbf{For}: $\tau = k+H-1,\ldots,k$ \\
															\hspace{1.05cm} \textbf{Step 4.1}: \textbf{Transmit}: $\mathbf{R}_i(\tau), \mathbf{B}_i(\tau)$ to $ \forall p \in \mathcal{D}_i^-\setminus \{i\}$;\\
															\hspace{4.05cm}$\mathbf{Q}_{p}(\tau+1)^{1/2}\mathbf{H}_{p,i}(\tau+1), \forall p\in \mathcal{D}_i^+$ to $\forall q\in \mathcal{D}_i^- \setminus \{i\}$;\\
															\hspace{4.05cm}\textbf{If:} $\tau \neq k$\\
															\hspace{5.05cm}$\mathbf{Q}_{i}(\tau)^{1/2}\mathbf{H}_{i,p}(\tau), \forall p\in \mathcal{D}_i^-$ to $\forall p\in \mathcal{D}_i^- \setminus \{i\}$;\\
															\hspace{4.05cm}\textbf{End if}\\
															\hspace{4.05cm}\textbf{If:} $\tau \neq k+H-1$\\
															\hspace{5.05cm}$\mathbf{R}_p(\tau+1), \mathbf{B}_p(\tau+1), \forall p\in \mathcal{D}_i^+$ to $\forall q\in \mathcal{D}_i^- \setminus \{i\}$;\\
															\hspace{5.05cm}$\mathbf{A}_i(\tau+1)$ to $\forall p\in \mathcal{D}_i^-$;\\
															\hspace{5.05cm}$\mathbf{K}_{p,i}(\tau+1), \forall p\in \mathcal{D}_i^+$ to $\forall q\in \mathcal{D}_i^- \setminus \{i\}$;\\
															\hspace{5.05cm}$\mathbf{P}_{i,(p,q)}(\tau+1)$ for some $(p,q)\in \phi_i$ to $\forall l\in \mathcal{D}_i^- \setminus \{i\}$.\\	
															\hspace{4.05cm}\textbf{End if}\\
															\hspace{1.05cm} \textbf{Step 4.2}: \textbf{Receive}: $\mathbf{R}_p(\tau), \mathbf{B}_p(\tau)$ from $ \forall p \in \mathcal{D}_i^+\setminus \{i\}$;\\
															\hspace{3.85cm}$\mathbf{Q}_{p}(\tau+1)^{1/2}\mathbf{H}_{r,p}(\tau+1), \forall r\in \mathcal{D}_i^+$ from $\forall p\in \mathcal{D}_i^+ \setminus \{i\}$;\\
															\hspace{3.85cm}\textbf{If:} $\tau \neq k$\\
															\hspace{4.85cm}$\mathbf{Q}_{p}(\tau)^{1/2}\mathbf{H}_{p,i}(\tau), \forall p\in \mathcal{D}_i^+$ from $p\in \mathcal{D}_i^+ \setminus \{i\}$;\\
															\hspace{3.85cm}\textbf{End if}\\
															\hspace{3.85cm}\textbf{If:} $\tau \neq k+H-1$\\
															\hspace{4.85cm}$\mathbf{R}_r(\tau+1), \mathbf{B}_r(\tau+1), \forall r\in \mathcal{D}_p^+$ from $\forall p\in \mathcal{D}_i^+ \setminus \{i\}$;\\
															\hspace{4.85cm}$\mathbf{A}_p(\tau+1)$ from $p\in \mathcal{D}_i^-$;\\
															\hspace{4.85cm}$\mathbf{K}_{r,p}(\tau+1), \forall r\in \mathcal{D}_p^+$ from $\forall p\in \mathcal{D}_i^+ \setminus \{i\}$;\\
															\hspace{4.85cm}$\mathbf{P}_{p,(r,s)}(\tau+1)$ for some $(r,s)\in \phi_p$ from $\forall p\in \mathcal{D}_i^+ \setminus \{i\}$.\\	
															\hspace{3.85cm}\textbf{End if}\\
															\hspace{1.05cm}\textbf{Step 4.3}: \textbf{Compute}:\\
															\hspace{2.45cm}\textbf{If:} $\tau = k +H-1$\\
															\hspace{3.45cm}$\mathbf{P}_{i,(p,q)}(\tau+1) \gets \sum_{r \in \mathcal{D}^+_p \cap \mathcal{D}^+_q}^{\phantom{a}}\mathbf{H}_{r,p}^T(\tau+1)\mathbf{Q}_{r}(\tau+1)\mathbf{H}_{r,q}(\tau+1), \forall (p,q)\in \phi_i$;\\
															\hspace{2.45cm}\textbf{Else}\\
															\hspace{3.45cm}Compute $\mathbf{P}_{i,(p,q)}(\tau+1)\forall (p,q)\in \phi_i$ making use of \eqref{eq:cov_prop_ass}.\\
															\hspace{2.45cm}\textbf{End if}\\
															\hspace{1.1cm}\textbf{Step 4.4}: \textbf{Compute}:\\
															\hspace{2.45cm}$\mathbf{S}_{p,q}(\tau) \gets \mathbf{B}_p^T(\tau)\mathbf{P}_{i,(p,q)}(\tau+1)\mathbf{B}_{q}(\tau) + \boldsymbol{\delta}_{pq}\mathbf{R}_q(\tau),  \forall (p,q)\in \phi_i$; \\
															\hspace{2.45cm}Compute $\tilde{\mathbf{S}}_i(\tau)$ and $\tilde{\mathbf{P}}_i(\tau+1)$ making use of \eqref{eq:def_Stilda} and \eqref{eq:def_Ptilda};\\
															\hspace{2.45cm}$\tilde{\mathbf{K}}_i(\tau) \gets \tilde{\mathbf{S}}_i(\tau)^{-1}\tilde{\mathbf{P}}_i(\tau+1)$;\\
															\hspace{1.1cm}\textbf{End for}\\
															\textbf{Step 5}: \textbf{Transmit}: $\mathbf{K}_{p,i}(\tau), \tau = k,\ldots,k+d-1$ to $\forall p\in \mathcal{D}_i^+\setminus\{i\}$.\\
															\textbf{Step 6}: \textbf{Receive}: $\mathbf{K}_{i,p}(\tau), \tau = k,\ldots,k+d-1$ from $\forall p\in \mathcal{D}_i^-\setminus\{i\}$.
														\end{algorithmic}
													\end{algorithm*}

													{Unsurprisingly, the inner loop of Algorithm~\ref{alg:OSDEKF} shares some similarities with the solution to the analogous distributed decentralized estimation problem recently proposed by the authors in \cite{pedroso2023distributed}. Although this work addresses a completely different problem with intricacies that far exceed the control-estimation duality, in \cite{pedroso2023distributed}[Remarks~2--4] noteworthy remarks are raised which are applicable to some implementation aspects of Algorithm~\ref{alg:OSDEKF}.}

													\subsection{Scheduling}\label{sec:scheduling_ti}
													
													In Section~\ref{subsec:synth_dec}, under appropriate approximations, an algorithm was devised for the decoupled synthesis of the RHC gains over a finite-window $\{k,\ldots, k+H-1\}$. This algorithm allows for distributing the global computation across the computational units of the systems that make up the network. Nevertheless, recall that, as put forward in Section~\ref{subsec:filter_def}, for the application of this framework to the infinite-horizon problem, a new window of gains of length $H$ has to be computed every $d$ time steps, of which only $d$ gains are used to compute the control input according to \eqref{eq:localFilter}. Furthermore, the decoupled gain computation in Algorithm~\ref{alg:OSDEKF} is carried out backwards in time. Thus, at the time instant that corresponds to the beginning of each window, all RHC gains over that window must have already been computed. Since these computations involve several communication instances, the steps in Algorithm~\ref{alg:OSDEKF} have to be properly scheduled to make use of soft real-time data transmissions only, as required by the communication constraints for a feasible large-scale implementation detailed in Section~\ref{subsec:com_req}. This issue is addressed in this section.
													
													
													
													Denote the control discretization time by $T_c$, which is the sampling time of the LTV system \eqref{eq:localDynamics}.  Let $T_t$ denote the interval of time between allowed communication instances, i.e., $T_t$ must be set so that it is greater than the time it takes for a system to communicate with the systems in its neighborhood and to perform floating-point operations with the data received in that transmission. Generally, $T_c$ is significantly larger than the minimum achievable $T_t$. It is important to remark that, in general, parallel to the control computations, there are also estimation algorithms running over the network. These usually require a higher rate of communication than control algorithms. Thus, one can set $T_t$ to the sampling time of the estimation algorithm.

													\begin{figure*}[ht]
														\includegraphics[width=0.85\linewidth]{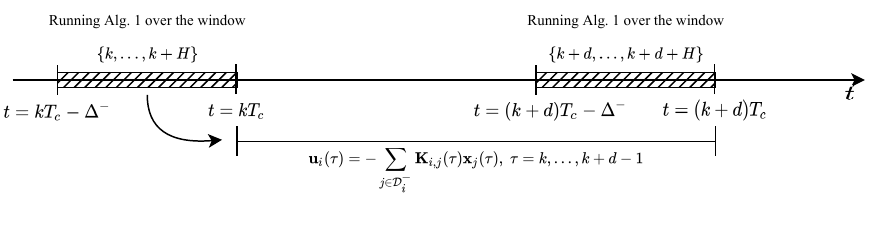}
														\centering
														\caption{Illustration of scheduling of Algorithm~\ref{alg:OSDEKF} over a timeline.}
														\label{fig:timeline_scheduling}
													\end{figure*}
													
													Algorithm~\ref{alg:OSDEKF} requires $H+2$ data transmissions between systems in the neighborhood of each other. Only after the last transmission of Algorithm~\ref{alg:OSDEKF}, the gains $\mathbf{K}_{i,p}(\tau), \forall p\in \mathcal{D}_i^-, \tau = k,\ldots,k+d-1$ are available to $\mathcal{T}_i$. To compute the control feedback according to \eqref{eq:localFilter}, these gains need to be available in $\mathcal{S}_i$ at $t = kT_c$. Therefore, to avoid hard real-time transmissions, they must be sent at most at $t = kT_c-T_t$. Thus, the proposed approach is to schedule the $H+2$ transmissions of Algorithm~\ref{alg:OSDEKF} such that they are performed at a rate of $1/T_t$. Thus, it takes $\Delta^- := (H+2)T_t$
													to run Algorithm~\ref{alg:OSDEKF}. For that reason, the computation of the RHC gains over the window $\{k,\ldots,k+H-1\}$ must start at $t = kT_c - \Delta^-$. A scheme of the proposed scheduling of Algorithm~\ref{alg:OSDEKF} over a timeline is depicted in Figure~\ref{fig:timeline_scheduling}{, for the illustrative time interval $[kT_c;(k+d)T_c[$. In this time interval, the control input is  computed according to \eqref{eq:localFilter}, making use of the gains in the window $\{k,\ldots,k+H-1\}$, which is represented in the line adjacent to the timeline that comprises $kT_c \leq t < (k+d)T_c$. As aforementioned, the computation of these gains must be carried out starting at $t = kT_c-\Delta^-$ and they only become available for use in the control law at $t = kT_c$, which is represented by the left rectangle with diagonal fill in the scheme. Moreover, during the application of the control inputs in $kT_c \leq t < (k+d)T_c$, the computation of the gains corresponding to the next finite window needs to start being carried out, which is represented in the scheme by the rectangle with diagonal fill in the right.}

													First, it is worth remarking that the proposed scheduling requires that the dynamics of a system $\mathcal{S}_i$ over the window $\{k,\ldots, k+H-1\}$ are predicted at $t = kT_c - \Delta^-$. Thus, if $\Delta^-$ is too large, the quality of the prediction may be degraded. Second, if $dT_c < \Delta^-$, i.e., $d/(H+2) < T_t/T_c$, the gain computation of two consecutive windows overlaps (rectangles with diagonal fill in Figure~\ref{fig:timeline_scheduling} overlap). If $T_t$ is not large enough to handle twice the communication and computational pressure, then it may lead to the unfeasibility of the control solution. 
													
													
													\subsection{Communication, computational, and memory requirements}\label{subsec:analysis_alg}
													
													
													To lighten the notation and enable a clearer analysis of the communication, computational, and memory requirements of Algorithm~\ref{alg:OSDEKF}, an homogeneous network is considered. Specifically, consider systems with the same order $n_1$, output dimension $o_1$, dimension of in-neighborhood $\nu^-_1$, and dimension of out-neighborhood $\nu^+_1$. 
													
													
													First, making use of the scheduling of the computations in Algorithm~\ref{alg:OSDEKF} proposed in Section~\ref{sec:scheduling_ti}, only soft real-time transmissions are required for the distributed computation of the controller gains. Second, the communication graph $\mathcal{G}_c$, {i.e.}, the graph representation of the required directed communication links, {corresponds to} the tracking output coupling graph with undirected edges. {Thus}, system $\mathcal{S}_i$ can only receive information from system $\mathcal{S}_j$ if the tracking output of $\mathcal{S}_j$ is coupled with the state of $\mathcal{S}_i$ or the converse, {i.e.}, $j\in \mathcal{D}^+_i \cup  \mathcal{D}^-_i$.  In fact, system $\mathcal{S}_i$ requires the exchange of data through communication with at most $\mathrm{max}\left(\nu_i^-,\nu_i^+\right)-1$ {systems at each iteration. The} communication complexity of system $\mathcal{S}_i$ is $\mathcal{O}\left(\mathrm{max}\left(\nu_i^-,\nu_i^+\right)\right)$. {Since neither $\nu_i^-$ nor $\nu_i^+$ increase with the number of systems in the whole network, the communication complexity of each system grows with $\mathcal{O}(1)$ with $N$.}
													
													
													Third, according to Algorithm~\ref{alg:OSDEKF}, each system $\mathcal{S}_i$ has to store in memory: i)~the matrices that model the dynamics of $\mathcal{S}_i$ over the window $\{k,\ldots,k+H-1\}$; ii)~the tracking output and control input weighting matrices of $\mathcal{S}_i$; iii)~$\mathbf{P}_{i,(p,q)}(\tau+1)$, with $(p,q)\in \phi_i$; and iv)~the sequences of controller gains $\mathbf{K}_{i,p}(\tau), \forall p\in \mathcal{D}_i^-, \tau = k,\ldots,k+d-1$ and $\mathbf{K}_{p,i}(\tau), \forall p\in \mathcal{D}_i^+, \tau = k,\ldots,k+H-1$. The data storage complexity of these is: i)~$\mathcal{O}\left(\mathrm{max}\left(n_1,m_1,\nu_1^-o_1\right)nH\right)$; ii)~$\mathcal{O}\left(\mathrm{max}\left(m_1^2,o_1^2\right)H\right)$; iii)~$\mathcal{O}({(\nu_1^+)}^2 n_1^2)$; and iv)~$\mathcal{O}(m_1n_1\mathrm{max}(\nu_1^+,\nu_1^-)H)$, respectively. {Even though some auxiliary variables are required to be handled at each iteration, their memory footprint is lower than the previously mentioned variables. Hence,} the data storage complexity of each system grows with $\mathcal{O}(1)$ with {$N$}.
													
													Fourth, the most intensive computations of Algorithm~\ref{alg:OSDEKF} are the propagation of $\mathbf{P}_{i,(p,q)}(\tau)$, with $(p,q)\in \phi_i$, in Step~4.3. These computations require $\mathcal{O}((\nu_1^-)^4n_1^3)$ floating-point operations in total, which grows with $\mathcal{O}(1)$ with {$N$}.
													
													
													{To conclude, Algorithm~\ref{alg:OSDEKF} abides by the communication, computational, and memory constraints detailed in Section~\ref{subsec:com_req}.} In \cite{pedroso2021discretecontrol} a decentralized RHC solution is also devised for networks of systems with LTV dynamics. However, using the method proposed therein, the global synthesis would have to be replicated in each computational unit, which violates Constraints~\ref{constraint:com}, \ref{constraint:storage}, and \ref{constraint:computational} {put forward} in Section~\ref{subsec:com_req}. {Therefore, the proposed solution makes great strides towards the implementation feasibility of decentralized control solutions over very large-scale networks.}
													
													
													\subsection{Extension to time-varying coupling topologies}\label{sec:tv_topology}
													
													Oftentimes, the tracking output couplings between systems vary with time due to: i)~the failure of systems of the network; ii)~the introduction of new systems in the network; or iii)~switching tracking configurations. The distributed and decentralized control solution is {now} extended to {accommodate} a time-varying tracking output coupling topology. {In that regard, consider} a time-varying directed graph $\mathcal{G}(k)$, with time-varying in-degree $\nu_i^-(k)$, in-neighborhood $\mathcal{D}^-_i(k)$, out-degree $\nu_i^+(k)$, out-neighborhood $\mathcal{D}^+_i(k)$, and define $\phi_i(k)$ and $\psi_i(k)$ analogously to \eqref{eq:def_phi} and \eqref{eq:def_psi}, respectively, for system $\mathcal{S}_i$.

													Algorithm~\ref{alg:OSDEKF_tv}, proposed in Appendix~\ref{app:alg2}, is the extension of Algorithm~\ref{alg:OSDEKF} to a time-varying tracking output coupling topology. Even though {such extension} is quite straightforward as far as distributing the gain synthesis across of the systems is concerned, that is not the case for the scheduling of the computations over time. Recall the scheduling procedure proposed in Section~\ref{sec:scheduling_ti}. It is possible to point out that the fact that the computation of the RHC gains over the window $\{k,\ldots,k+H-1\}$ must start at $t = kT_c - \Delta^-$, requires that at $t = kT_c - \Delta^-$ a system $\mathcal{S}_i$ receives data through communication from $\mathcal{S}_p$, with $p \in \mathcal{D}^+_i(k+H)$. That is, at $t = kT_c - \Delta^-$ communication between $\mathcal{S}_i$ and a system with which $\mathcal{S}_i$ is coupled at $t = (k+H)T_c$ is required. In most applications, if two systems are coupled at a given time instant, communication between them is probably feasible at that time instant. However, at a given time instant, communication between two systems that are coupled an interval of time $HT_c+\Delta^-$ may later not be feasible due to the changing spatial configuration of the network over time. For example, consider the problem of maintaining a formation of unnamed aerial vehicles (UAVs). At a given time instant, two UAVs may be separated by an obstacle. If a tracking output coupling is expected to be established between them in the future and if $H$ is required to be large, then communication between them may be required when they are still separated by the obstacle, which may be difficult to achieve. It is important to point out that the effect of this aspect varies greatly with the application in question. On one hand, in applications with slowly time-varying coupling topologies and spatial configurations, the communication requirements needed to implement the scheduling procedure proposed in Section~\ref{sec:scheduling_ti} are likely feasible. On the other hand, in applications that, in the time frame of the receding horizon window, there are significant spatial configuration changes that impede communication links between systems, then some additional considerations should be taken into account.

													\section{Application to onboard orbit control of LEO mega-constellations}\label{sec:app_sats}
													
													In this section, the distributed decentralized RHC algorithm developed in Section~\ref{sec:DEKF} is applied to the cooperative orbit control problem of LEO mega-constellations. The scheme presented in this section is novel and it is developed aiming for efficiency and fuel saving in a distributed and decentralized framework. {Unlike other approaches in the literature, it follows the crucial constraints detailed in Section~\ref{subsec:com_req} for a feasible real-time implementation in a very large-scale. The source code of the numerical simulations is available in the DECENTER Toolbox at} {\small \url{https://decenter2021.github.io/examples/DDRHCStarlink/}}.

													\subsection{Mega-constellation model} \label{sec:mega_leo_model}
													
													{Consider a Walker constellation $\bar{i}: T/P/F$. It has a total of $T$ satellites, evenly distributed over $P$ orbital planes at a nominal inclination $\bar{i}$, whose} nominal relative phasing between adjacent planes is $\bar{\beta} = 2\pi F/T$, where $F$ is the phasing parameter. The nominal orbits are circular and have a semi-major axis of $\bar{a}$. This constellation can be modeled as a network of coupled systems, $\mathcal{S}_j$, each associated with a computational unit $\mathcal{T}_j$, with $j = 1,...,T$. Let $\mathbf{p}_i \in \mathbb{R}^3$ and $\mathbf{v}_i \in \mathbb{R}^3$ denote the position and velocity vectors, respectively, of $\mathcal{S}_i$ expressed in the J2000 Earth centered inertial (ECI) frame. The dynamics of each satellite of the constellation are modeled independently, since there is no dynamical coupling between them. Each satellite $\mathcal{S}_i$ is equipped with Hall effect thrusters, aligned according to the local TNW frame ($x$ axis along the velocity vector, $z$ axis along the orbit's angular momentum vector, and $y$ axis completes the right-handed coordinate system) that generate a force $\mathbf{u}_i \in \mathbb{R}^3$ expressed in the TNW local frame. Each thruster has a maximum thrust, $C_{t1}$. The model of the dynamics of a single satellite $\mathcal{S}_i$ is, thus, given by
													\begin{equation}\label{eq:initial_nn_dyn}
														\begin{cases}
															\dot{\mathbf{p}}_i = \mathbf{v}_i\\
															\dot{\mathbf{v}}_i = -\mu{\mathbf{p}_i}/{||\mathbf{p}_i||_2^3} + \mathbf{a}^{J_2}_i \!+ \mathbf{a}^P_i \!+ \mathbf{R}_{TNW}^{ECI}\mathbf{u}_i/{m_i}\!\!\\
															\dot{m}_i = -||\mathbf{u}_i||_1/(I^{sp}_i g_0),
														\end{cases}
													\end{equation}
													where $m_i$ denotes the mass of the satellite, $\mu$ denotes the gravitational parameter of the Earth, $\mathbf{a}^{J_2}_i \in \mathbb{R}^3$ and $\mathbf{a}^P_i \in \mathbb{R}^3$ denote the perturbation accelerations of the effect of $J_2$ and all other perturbations, respectively, $\mathbf{R}_{TNW}^{ECI}$ denotes the rotation matrix from the TNW local frame to the J2000 ECI frame,  $I^{sp}_i$ denotes the specific impulse of the Hall effect thrusters of $\mathcal{S}_i$, and $g_0$ denotes the standard gravity acceleration.
													
													In this application, for control law synthesis purposes, the parameterization of the orbits of each satellite of the constellation is achieved by the set of non-singular mean orbital elements for near-circular inclined orbits $(a,u,e_x,e_y,i,\Omega)$, respectively semi-major axis, mean argument of latitude, two eccentricity vector components, inclination, and longitude of ascending node. These can be related with the more common Keplerian set $(a,e,i,\Omega,\omega, M)$, which has a singularity for circular orbits, with 
													\begin{equation*}
														\begin{cases}
															u = M+ \omega\\
															e_x = e\cos \omega\\
															e_y = e \sin \omega,
														\end{cases}
													\end{equation*} 
													where $e$ denotes the eccentricity, $\omega$ denotes the argument of perigee, and $M$ denotes the mean anomaly. Denote the state of a satellite $\mathcal{S}_i$, made up of the aforementioned six non-singular mean orbital elements, by 
													\begin{equation*}
														\mathbf{x}_i(t) = \left[ a_i(t) \;  u_i(t) \;  e_{xi}(t) \;  e_{yi}(t) \;  i_i(t) \; \Omega_i(t) \right]^T.
													\end{equation*}
													
													The satellite orbital mechanics \eqref{eq:initial_nn_dyn} are nonlinear and, thus, have to be linearized to employ the distributed and decentralized method put forward in this paper. The linearization of the dynamics of each satellite is carried out about a nominal orbit. These are defined such that the set of nominal orbits of all satellites makes up a consistent nominal constellation, in the sense that the nominal separations: i)~along-track; ii)~inter-plane; and iii)~ in relative phasing between adjacent planes are enforced. It is very important to remark that this nominal constellation is used for linearization purposes only -- it is not employed for bounding-box tracking of each individual satellite at any point. The necessity of enforcing constellation-wide constraints in the definition of each nominal orbit is made clear in the formulation of the orbit control problem as a RHC problem in Section~\ref{subsec:app_ctrl}. 
													
													The nominal state of $\mathcal{S}_i$ at time instant $t$, $\bar{\mathbf{x}}_i(t) = \left[ \bar{a}_i(t) \;  \bar{u}_i(t) \;  \bar{e}_{xi}(t) \;  \bar{e}_{yi}(t) \;  \bar{i}_i(t) \; \bar{\Omega}_i(t) \right]^T$, is defined by
													\begin{equation}\label{eq:nominal_orbit}
														\begin{cases}
															\bar{a}_i(t) = \bar{a}\\
															\bar{u}_i(t) =  \bar{u}_{t_0} + \left( (i-1) \!\!\!\!\mod T/P\right)2 \pi P/T \\
															\quad  + \lfloor (i\!-\!1)P/T \rfloor 2 \pi F/T + (\dot{M}+\dot{\omega})(t-t_0)\\
															\bar{e}_{x,i}(t) = 0\\
															\bar{e}_{y,i}(t) = 0\\
															\bar{i}_i(t) = \bar{i}\\
															\bar{\Omega}_i(t) = \bar{\Omega}_{t_0} + \lfloor (i\!-\!1)P/T \rfloor 2 \pi /P + \dot{\Omega}(t-t_0).\!\!
														\end{cases}
													\end{equation}
													Above, $\dot{M}$, $\dot{\omega}$, and $\dot{\Omega}$ are the secular rates, including the effect of $J_2$, on the mean anomaly, argument of perigee, and longitude of ascending node, respectively, which are given by \cite[Chapter 8]{vallado}
													\begin{equation*}
														\dot{M} = \frac{3\bar{n}R_{\bigoplus}^2J_2}{4\bar{a}^2}\left(3\sin^2\bar{i}-2\right),
													\end{equation*}
													\begin{equation*}
														\dot{\omega} = \frac{3\bar{n}R_{\bigoplus}^2J_2}{4\bar{a}^2}\left(4-5\sin^2\bar{i}\right),
													\end{equation*}
													and
													\begin{equation*}
														\dot{\Omega} = -\frac{3\bar{n}R_{\bigoplus}^2J_2}{2\bar{a}^2}\cos \bar{i},
													\end{equation*}
													for a nominal orbit, where $\bar{n} := \sqrt{\mu/\bar{a}^3}$, and $R_{\bigoplus}$ denotes the Earth's mean equatorial radius. Note that the nominal orbits of all satellites in \eqref{eq:nominal_orbit} depend on three constellation-wise parameters $(t_0,\bar{u}_{t_0},\bar{\Omega}_{t_0})$, whose physical meaning is that the nominal orbit of $\mathcal{S}_1$ has mean argument of latitude $\bar{u}_{t_0}$ and longitude of ascending node $\bar{\Omega}_{t_0}$ at time instant $t_0$. These three parameters are designed herein as the anchor of the nominal constellation. 
													
													There are a few aspects worth pointing out regarding the anchor of the nominal constellation. First, all satellites must agree on an anchor for the nominal constellation at any time instant. Second, to minimize linearization errors, the anchor should be selected such that the nominal position of each satellite is as close as possible to their actual position. Thus, since the position of the satellites drifts away from their nominal position with time, due to neglected secular effects, other perturbations, and maneuvers, the anchor must be updated from time to time. Third, note that the evolution of the nominal states takes the effect of the Earth's oblateness into account, which significantly decreases the frequency with which the anchor has to be updated. Fourth, the computation of the anchor for a time instant $t_0$ should be performed in accordance with an optimization problem of the form
													\begin{equation*}
														\underset{\bar{u}_{t_0},\bar{\Omega}_{t_0}}{\text{minimize}} \sum_{i = 1}^{T}\left( \alpha\left( u_i(t_0)-\bar{u}_i(t_0) \right) + \alpha\left( \Omega_i(t_0)-\bar{\Omega}_i(t_0) \right) \right),
													\end{equation*}
													where $\alpha : \mathbb{R} \to  \mathbb{R}$ is a convex function. Given that $\bar{u}_i(t_0)$ does not depend on $\bar{\Omega}_{t_0}$ and $\bar{\Omega}_i(t_0)$ does not depend on $\bar{u}_{t_0}$, the optimization above can be decoupled into two problems: one corresponding to $\bar{u}_{t_0}$ and other to $\bar{\Omega}_{t_0}$. Fifth, for the sake of simplicity, in this application, $\alpha(\cdot) = (\cdot )^2$ is chosen, which leads to the closed-form solution 
													\begin{equation}\label{eq:anchor_computation}
														\begin{cases}
															\bar{u}_{t_0} \!= \!\sum_{i=1}^{T} \left(u_i(t_0) \!-\!\left( (i\!-\!1) \!\!\!\!\mod T/P\right)2 \pi P/T \right. \!\!\!\!
															\\ \quad \quad \quad \quad \quad \quad \quad \quad \:\,- \left.  \lfloor (i-1)P/T \rfloor 2 \pi F/T\right)\\
															\bar{\Omega}_{t_0} = \sum_{i=1}^{T} \left(\Omega_{i}(t_0) - \lfloor (i-1)P/T \rfloor 2 \pi /P \right).
														\end{cases}
													\end{equation}
													However, to improve the robustness to outliers, the $\ell_1$ norm, i.e., $\alpha(\cdot) = |\cdot|$, or Huber loss function could be used instead, still leading to a convex optimization problem. Sixth, even though the optimization problem above cannot be easily decoupled to be distributed across the computational units of the satellites in the network, it is not a serious issue, since it is only required to be solved sporadically. Thus, it can either be: i)~computed in a centralized node and then the solution broadcast to the network; or ii)~solved distributively over a period of time making use of distributed gradient methods with asymptotic consensus guarantees \cite{jakovetic2014fast}. It is worth pointing out that a new anchor $(t_0,\bar{u}_{t_0},\bar{\Omega}_{t_0})$ can be used starting at $t = t_0 + \Delta$, where $\Delta$ can be as large as necessary to allow for communication and centralized or cooperative computations.
													
													The evolution of the state of $\mathcal{S}_i$ is linearized about the aforementioned nominal orbits, defining a relative position $\delta \mathbf{x}_i(t)$ based the set of orbital elements $\delta \mathbf{x}_i(t) := [a_i(t)\;\delta u_i(t) \; \delta e_{x,i}(t) \; \delta e_{y,i}(t) \; \delta i_i(t) \; \delta \Omega_i(t)]$,  introduced in \cite{d2010autonomous}, which is defined as
														\begin{equation}\label{eq:relative_mean_oe_trans}
															\delta \mathbf{x}_i(t) := \begin{bmatrix}
																a_i(t)/\bar{a}_i(t)-1\\
																u_i(t)-\bar{u}_i(t) + \left(\Omega_i(t)-\bar{\Omega}_i(t)\right)\cos \bar{i}_i(t)\\
																e_{x,i}(t)-\bar{e}_{x,i}(t)\\
																e_{y,i}(t)-\bar{e}_{y,i}(t)\\
																i_i(t)-\bar{i}_i(t)\\
																\left(\Omega_i(t)-\bar{\Omega}_i(t)\right)\sin \bar{i}_i(t)
															\end{bmatrix}.
														\end{equation}
														This set parameterizes the position of the satellite, $\mathbf{x}_i(t)$, in relation to its nominal position, $\bar{\mathbf{x}}_i(t)$. In \cite{sullivan2016improved,koenig2016new} and \cite{di2018continuous}, the dynamics \eqref{eq:initial_nn_dyn}, taking the effect of $J_2$ into account but neglecting the remaining perturbations, are linearized about near-circular nominal orbits. Making use of Floquet theory, system transition and convolution matrices are derived to write the discrete-time LTV system 
														\begin{equation}\label{eq:lin_rel_dyn_sat}
															\delta \mathbf{x}_i((k+1)T_c) =  \mathbf{A}_i(k)\delta\mathbf{x}_i(kT_c)+ \mathbf{B}_i(k) \mathbf{u}_i(kT_c)/m_i(kT_c) 
														\end{equation}
														with a sampling time $T_c$ and assuming that $\mathbf{u}_i(t)$ and $m_i(t)$ remain constant over each interval $\left[kT_c;(k+1)T_c\right[$. Given that electric propulsion is employed, which has very reduced propellant mass rates, the constant mass approximation is very reasonable (as an example, the Hall effect thrusters considered in the illustrative simulations in the sequel reach a mass rate of the order of $10^{-6}$ Kg/s at full throttle). Henceforth, to alleviate the notation, the continuous time instant $t = kT_c$ is denoted by the discrete-time index $k$. As an example, $\delta \mathbf{x}_i(kT_c)$ and $\mathbf{u}_i(kT_c)$ are denoted by $\delta \mathbf{x}_i(k)$ and  $\mathbf{u}_i(k)$, respectively. For circular nominal orbits and following the notation herein, the state transition matrix $\mathbf{A}_i(k)$ is given, according to \cite{di2018continuous}, by
														\begin{equation*}
															\begin{split}
																\mathbf{A}_i(k) = \begin{bmatrix}
																	1 & 0 & 0 & 0 & 0 & 0\\
																	-\bar{\Lambda} T_c & 1 & 0 & 0 & \! -4\bar{K} T_c \sin 2\bar{i} \! & 0 \\ 
																	0 & 0  & \!\cos \Delta \omega \! & \!- \sin \Delta \omega \! & 0 & 0 \\
																	0 & 0  & \!\sin \Delta \omega \! &  \!\cos \Delta \omega \! & 0 & 0 \\
																	0 & 0 & 0 & 0 & 1 & 0\\
																	\frac{7}{2}\bar{K}T_c\sin 2\bar{i} \! & 0 & 0 & 0 & 2\bar{K}T_c\sin^2 \bar{i} & 1\\
																\end{bmatrix}
															\end{split}
														\end{equation*}
														and the convolution matrix $\mathbf{B}_i(k)$ by
														\begin{equation}\label{eq:def_B}
															\mathbf{B}_i(k) = \begin{bmatrix}
																\frac{2 \Delta u}{\bar{n}\bar{a}\bar{W}} & 0 & 0 \\
																-\frac{\bar{\Lambda} \Delta u^2}{\bar{n}\bar{a}\bar{W}^2} & \frac{2 \Delta u}{\bar{n}\bar{a}\bar{W}} & \Psi_{2,3} \\
																2 \Psi_{4,2} &  \Psi_{3,2} & 0\\
																-2 \Psi_{3,2} &  \Psi_{4,2} & 0\\
																0 & 0 & \Psi_{5,3} \\
																\frac{7}{2}\frac{\bar{K}\Delta u^2\sin 2\bar{i}}{\bar{n}\bar{a}\bar{W}^2} & 0 & \Psi_{6,3}
															\end{bmatrix},
														\end{equation}
														where
														\begin{align*}
															&\Psi_{2,3} := \frac{4\bar{K}\sin 2\bar{i}}{\bar{n}\bar{a}\bar{W}^2} \; \Big(\!\cos (\bar{u}_i(kT_c+T_c))-\cos(\bar{u}_i(kT_c))  \\
															&\quad \quad \quad \quad \quad \quad \quad \quad +\sin(\bar{u}_i(kT_c))\Delta u\Big),\\
															& \Psi_{3,2} :=  \frac{\cos(\bar{u}_i(kT_c+T_c))-\cos(\bar{u}_i(kT_c)+\bar{C}\Delta u)}{\bar{n}\bar{a}(1-\bar{C})\bar{W}},\\
															& \Psi_{4,2} := \frac{\sin(\bar{u}_i(kT_c+T_c))-\sin(\bar{u}_i(kT_c)+\bar{C}\Delta u)}{\bar{n}\bar{a}(1-\bar{C})\bar{W}},\\
															&\Psi_{5,3} := \frac{\sin(\bar{u}_i(kT_c+T_c))-\sin(\bar{u}_i(kT_c))}{\bar{n}\bar{a}\bar{W}},\\
															&\Psi_{6,3} \!:= - \frac{\bar{W}\!+\!2\bar{K}\sin^2 \bar{i}}{\bar{n}\bar{a}\bar{W}^2} (\cos (\bar{u}_i(kT_c\!+\!T_c))-\cos(\bar{u}_i(kT_c)))\\
															&\quad \quad \quad \,- {2\bar{K}\sin^2 \bar{i}\sin(\bar{u}_i(kT_c))\Delta u}/\left(\bar{n}\bar{a}\bar{W}^2\right),\\
															&\Delta u := \bar{W}T_c,\\
															&\Delta \omega := \bar{K}(5\cos^2 \bar{i}-1)T_c,\\
															&\bar{K} := (3/4)\bar{n}R_{\bigoplus}^2J_2/\bar{a}^2,\\
															&\bar{\Lambda} := (3/2)\bar{n} + (7/2)\bar{K}(3\cos^2 \bar{i}-1),\\
															&\bar{W} := \bar{n} + \bar{K}(8\cos^2 \bar{i}-2), \quad \text{and}\\
															&\bar{C} := \bar{K}(5\cos^2 \bar{i}-1)/\bar{W}.
														\end{align*}
														Note that $\mathbf{A}_i(k)$ is time-invariant and equal for all satellites, whereas $\mathbf{B}_i(k)$ is time-varying and depends on the known nominal evolution of the mean argument of latitude. Thus, these matrices can be easily predicted in $\mathcal{T}_i$ over a window of future time instants.
														
														\subsection{Controller implementation}\label{subsec:app_ctrl}
														
														Now that the constellation has been modeled as a LTV system, the constellation orbit control problem has to be formulated as a RHC problem. A common approach is the bounding box method. In this scheme, a reference position is generated for each satellite around which an error box is defined. Whenever each satellite is inside the error box no control input is used, but when it leaves said box the feedback control is enabled to drive it inside of the error box. The main advantage of this scheme is that the low-level control feedback loop of each satellite is decoupled from the others. However, this decoupled scheme tries to correct common secular and periodic perturbations that cause the satellites to drift in relation to the nominal constellation but that perturb the constellation shape to a much lesser extent, thus wasting much-valuable fuel. If, to mitigate this effect, the nominal positions are updated in real-time, then the global computation of consistent nominal positions for each satellite has to be carried out in a centralized node or cooperatively across the network in real-time. Nevertheless, this alternative requires tremendous communication load, which is unfeasible for large-scale networks.
														
														In this paper, in an attempt to reduce fuel consumption and to follow the communication, computational, and memory constraints detailed in Section~\ref{subsec:com_req}, a control scheme is devised such that the satellites control their position relative to each other. On one hand, the semi-major axis, eccentricity, and inclination of the orbit of each satellite may be controlled in a decoupled fashion, thus an inertial tracking output component given by
														\begin{equation*}
															\mathbf{z}_{i,in}(k) = \begin{bmatrix}
																a_i(k)-\bar{a}_i(k)\\
																e_{x,i}(k)-\bar{e}_{x,i}(k)\\
																e_{y,i}(k)-\bar{e}_{y,i}(k)\\
																i_i(k)-\bar{i}_i(k)
															\end{bmatrix} = \begin{bmatrix}
																\bar{a}_i(k)\delta a_i(k) \\
																\delta e_{x,i}(k) \\
																\delta e_{y,i}(k) \\
																\delta i_i(k)
															\end{bmatrix}
														\end{equation*}
														is considered for each satellite $\mathcal{S}_i$, which is not coupled with any other satellites. Note that driving $\mathbf{z}_{i,in}(k)$ to zero is equivalent to driving the semi-major axis, eccentricity, and inclination to their nominal values. On the other hand, to maintain the shape of the constellation, $\delta u_i(k)$ and $\delta \Omega_i(k)$ ought to be controlled in relation to other satellites. To achieve a distributed solution, each satellite ought to be coupled with only a small number of satellites, which does not scale with the number of satellites in the constellation. Furthermore, the satellites with which $\mathcal{S}_i$ is coupled should be in its proximity, which is more convenient to establish communication links. Thus, it is considered that two satellites are coupled if they are within a tracking range $R$ of each other, i.e., $||\mathbf{p}_i-\mathbf{p}_j|| \leq R$, up to a maximum of $|\mathcal{D}^-|_{\text{max}}$ satellites in $\mathcal{D}^-_i$. If more than $|\mathcal{D}^-|_{\text{max}}-1$ satellites other than $\mathcal{S}_i$ are within a tracking range of $\mathcal{S}_i$, only the $|\mathcal{D}^-|_{\text{max}}-1$ closest are considered. Since the nominal evolution of the constellation is known, it is easy to predict the coupling topology over a window of future time instants. Let $\mathcal{D}^-_i \setminus \{i\} = \left\{j^i_1, \ldots,j^i_{|\mathcal{D}^-_i|-1}\right\}$. Then the relative tracking output component is given by 
														\begin{equation*}
															\mathbf{z}_{i,rel}(k) := \mathrm{col}\left(\mathbf{z}_{i,j^i_1}^{ref}(k),\ldots, \mathbf{z}_{i,j^i_{|\mathcal{D}^-_i|-1}}^{ref}(k) \right)
														\end{equation*}
														with 
														\begin{equation}\label{eq:output_rel}
															\begin{split}
																\mathbf{z}_{i,j}^{ref}(k) :=  &\begin{bmatrix}
																	u_i(k)-u_j(k)-\left(\bar{u}_i(k)-\bar{u}_j(k)\right)\\
																	\Omega_i(k)-\Omega_j(k)-\left(\bar{\Omega}_i(k)-\bar{\Omega}_j(k)\right)\\
																\end{bmatrix} \\
																=&\begin{bmatrix}
																	\delta u_i(k)\!-\!\delta u_j(k) \!-\!\left(\delta \Omega_i(k) \!-\!\delta \Omega_j(k) \right)/ \tan \bar{i}\\
																	\left(\delta \Omega_i(k)-\delta \Omega_j(k)\right)/ \sin \bar{i}
																\end{bmatrix}.
															\end{split}
														\end{equation}
														Thus, if $\mathcal{G}(k)$ contains a directed spanning tree, driving $\mathbf{z}_{i,j}^{ref}(k)$ to zero maintains the shape of the constellation. It is interesting to point out that the definition of the relative tracking output \eqref{eq:output_rel} makes use of the nominal constellation just to retrieve the nominal angular spacing in $u$ and $\Omega$ between $\mathcal{S}_i$ and $\mathcal{S}_j$. Thus, the actual position of the satellite is inevitably going to slowly drift away from the nominal constellation, while maintaining the desired shape.  Defining the tracking output of $\mathcal{S}_i$ as $\mathbf{z}_i(k) := \mathrm{col}(\mathbf{z}_{i,rel}(k),\mathbf{z}_{i,in}(k))$, it can be written as 
														\begin{equation}\label{eq:app_ouput}
															\mathbf{z}_i(k) = \sum_{p\in \mathcal{D}^-_i}\mathbf{H}_{i,p}(k)\delta \mathbf{x}_p(k)
														\end{equation}
														with
														\begin{equation*}
															\mathbf{H}_{i,p}(k) := \begin{cases}
																\begin{bmatrix}\mathbf{1}_{|\mathcal{D}^-_i|\times 1 }  \otimes \mathbf{H}_{rel}\\\mathbf{H}_{in}\end{bmatrix},&\;p = i\\
																\vspace{-0.3cm}\\
																-\begin{bmatrix}\mathbf{l}_{k} \otimes  \mathbf{H}_{rel} \\\mathbf{0}_{4\times 6}\end{bmatrix} ,&\;p = j_{i,k},
															\end{cases}
														\end{equation*}
														where
														\begin{equation*}
															\mathbf{H}_{in}:= \begin{bmatrix}
																\bar{a} & 0 & 0 & 0 & 0 & 0\\
																0 & 0 & 1 & 0 & 0 & 0\\
																0 & 0 & 0 & 1 & 0 & 0\\
																0 & 0 & 0 & 0 & 1 & 0\\
															\end{bmatrix},
														\end{equation*}
														and
														\begin{equation*}
															\mathbf{H}_{rel}:= \begin{bmatrix}
																0 & 1 & 0 & 0 & 0 & -1/\tan \bar{i}\\
																0 & 0 & 0 & 0 & 0 & 1/\sin \bar{i}
															\end{bmatrix}.
														\end{equation*}
														The tracking output weighting matrices $\mathbf{Q}_i(k)$ are of the form 
														\begin{equation*}
															\mathbf{Q}_i(k) := \mathrm{diag}\left(\mathbf{I}_{|\mathcal{D}_i^-|-1} \otimes \mathbf{Q}_i^{rel}(k),\mathbf{Q}_i^{in}(k)\right),
														\end{equation*}
														where $\mathbf{Q}_i^{rel}(k) \in \mathbb{R}^{2\times 2}$, $\mathbf{Q}_i^{in}(k) \in \mathbb{R}^{4\times 4}$, and $\mathbf{R}_i(k) \in \mathbb{R}^{3\times 3}$ are defined in the sequel for the illustrative constellation under study. 
														
														
														
														The evolution of the state of each satellite state is modeled by the LTV system \eqref{eq:lin_rel_dyn_sat} and the constellation orbit control problem is formulated as the regulation of the tracking output \eqref{eq:app_ouput}. Thus, we are in the conditions of applying the distributed and decentralized RHC method put forward in Section~\ref{sec:DEKF}, which  abides by the communication, computational, and memory feasibility constraints detailed in Section~\ref{subsec:com_req}. The tracking output coupling graph $\mathcal{G}(k)$  is time-varying, thus one has to follow Algorithm~\ref{alg:OSDEKF_tv}. Considering the convolution matrix as defined in \eqref{eq:def_B}, one obtains the feedback law
														\begin{equation*}\label{eq:localFilter_app}
															\mathbf{u}_i(k) = - m_i(k)\sum_{j\in \mathcal{D}^-_i(k)}\mathbf{K}_{i,j}(k)\delta\mathbf{x}_j(k).
														\end{equation*}
														However, more often than not, the relative mean orbital elements are not readily available to each satellite from onboard sensors or filters. It is more usual for each satellite to have access to its position and velocity in a Cartesian coordinate system, which can be obtained making use of a GNSS receiver, for instance. Although it is easy to compute the relative mean orbital elements from the mean orbital elements and nominal orbital elements according to \eqref{eq:relative_mean_oe_trans}, it is not straightforward to obtain the mean orbital elements from the Cartesian position and velocity. In fact, if ones applies the Keplerian orbit transformation directly to the Cartesian position and velocity, one obtains osculating orbital elements, which contain very significant short-period oscillations due to the Earth's uneven gravity field and other perturbations. For instance, in the conditions of the illustrative simulations in the sequel, these oscillations lead to differences between osculating and mean orbital elements that reach 6 km in the semi-major axis. If one neglected these differences, the controller feedback would attempt to counteract these natural oscillations, thus wasting fuel and degrading tracking performance. To aim for meter-level tracking accuracy and to reduce fuel consumption, a transformation inspired in the one proposed in \cite{spiridonova2014precise} is employed to account for the Earth's uneven gravity field making use of the spherical harmonic expansion up to a desired degree, which is the main source of these oscillations. An open-source MATLAB implementation and thorough documentation of this transformation is available in the \textit{osculating2mean} toolbox at {\small \url{https://github.com/decenter2021/osculating2mean}}, which is based on \cite{eckstein1970reliable,kaula2013theory,hwang2001gravity} and \cite{hwang2002satellite}. In the simulation results in the sequel, the Earth’s gravity field EGM96 spherical harmonic expansion \cite{lemoine1998development} up to degree and order 12 is employed.

														\subsection{Illustrative mega-constellation and tuning}\label{sec:simulations}
														
														The illustrative mega-constellation of a single shell, inspired in the first shell of the Starlink constellation, was chosen to assess the performance of the method devised in this paper. In this section, the illustrative mega-constellation is described and the parameters of the control solution are tuned. The constellation is a Walker $53.0\deg:1584/72/17$. The phasing parameter of this Starlink shell is {chosen so that the minimum distance between two satellites is maximized \cite{liang2021phasing}. Table~\ref{tab:constellation_parameters} presents the parameters that characterize this shell. All satellites are assumed to be identical.}
														
														\begin{table}[ht!]
															\centering
															\caption{Parameters of the constellation.}
															\label{tab:constellation_parameters}
															\begin{tabular}{lc}
																\hline
																Configuration & \\
																\hline
																Inclination ($\bar{i}$) & $53.0\deg$\\
																Number of satellites ($T$) & 1584\\
																Number of orbital planes ($P$) & 72\\
																Phasing parameter ($F$) & 17\\
																Semi-major axis ($\bar{a}$) &  $6921.0$ km \\
																Eccentricity ($\bar{e}$) & 0\\
																\hline 
																\hline
																\\[-1em]
																Satellites & \\
																\hline
																\\[-1em]
																Initial mass  & $260$ Kg\\
																Drag coefficient ($C_D$) & 2.2\\
																Section area ($A$) & $24.0$ m$^2$\\
																Solar radiation pressure coefficient ($C_R$) & 1.2\\
																Solar radiation pressure area ($\mathrm{SRPA}$) & $10.0$ m$^2$\\
																\hline 
																\hline
																\\[-1em]
																Electric thrusters & \\
																\hline
																\\[-1em]
																Maximum thrust ($C_{t1}$) & 0.068 N\\
																Specific impulse ($I_{sp}$) & 1640.0 s\\
																\hline
															\end{tabular}
														\end{table}

\begin{figure}[]
															\centering
															\includegraphics[width=\linewidth]{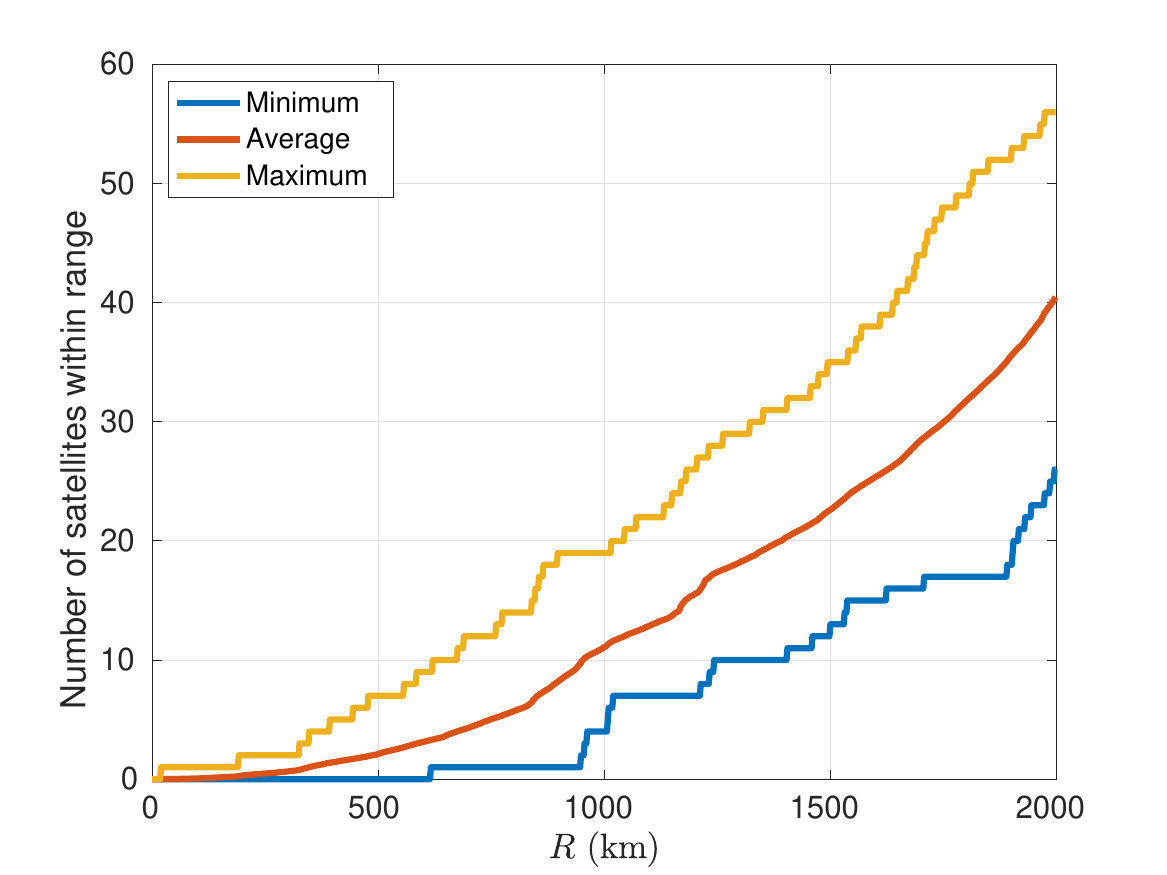}
															\caption{Number of satellites within ISL range at 0 TDB seconds since J2000.}
															\label{fig:isl_range}
														\end{figure}

\begin{figure*}[ht!]
															\centering
															\includegraphics[width=\textwidth]{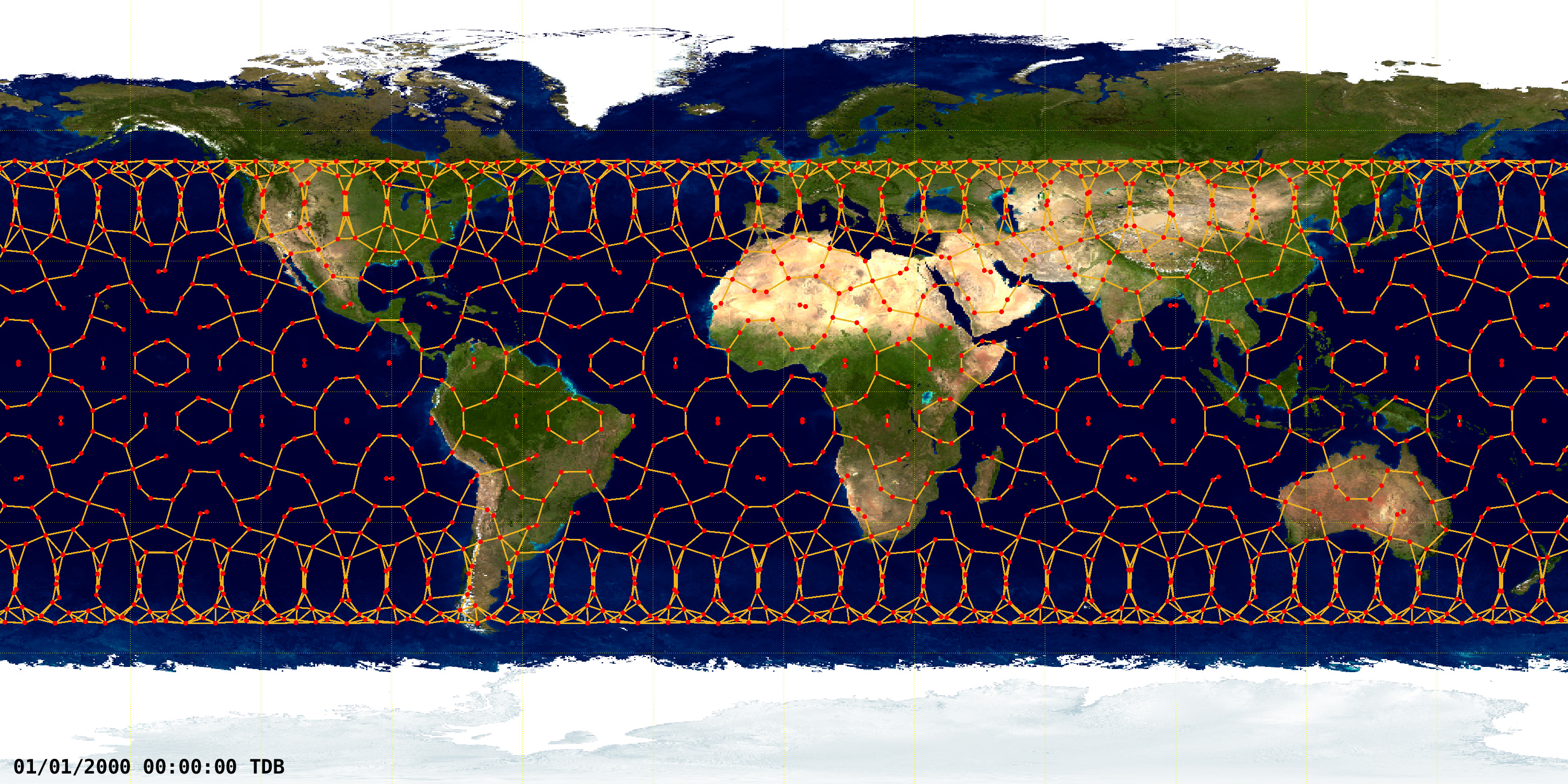}
															\caption{Snapshot of ground track and ISL of the simulated constellation at 0 TDB seconds since J2000.}
															\label{fig:ground_track}
														\end{figure*}

														{Figure~\ref{fig:isl_range} depicts the minimum, maximum, and average number of satellites within tracking output coupling range among all satellites as a function of $R$, at 0 Dynamical Barycentric Time (TDB) seconds since J2000.} A tracking output coupling range of $R =750$ km is considered, {which enables all satellites to establish} a tracking output coupling with, at least, another satellite at any time. The maximum in-neighborhood cardinality is set to  $|\mathcal{D}^-|_{\text{max}} = 6$, which allows for each satellite to establish tracking output couplings with up to $5$ other satellites. In Figure~\ref{fig:ground_track}, a snapshot of the projection of the position of each satellite of the constellation over the Earth's surface, as well as the tracking output couplings, at 0 TDB seconds since J2000 is shown. An animation of the evolution of the ground track of the constellation and of the tracking output couplings can be viewed in the website of the DECENTER Toolbox. It is interesting to note that, due to the higher density of satellites at extreme latitudes, much more couplings are established. This fact allows for more accurate control in these regions, which is desirable to avoid collisions.

\begin{figure}[]
															\centering
															\includegraphics[width=\linewidth]{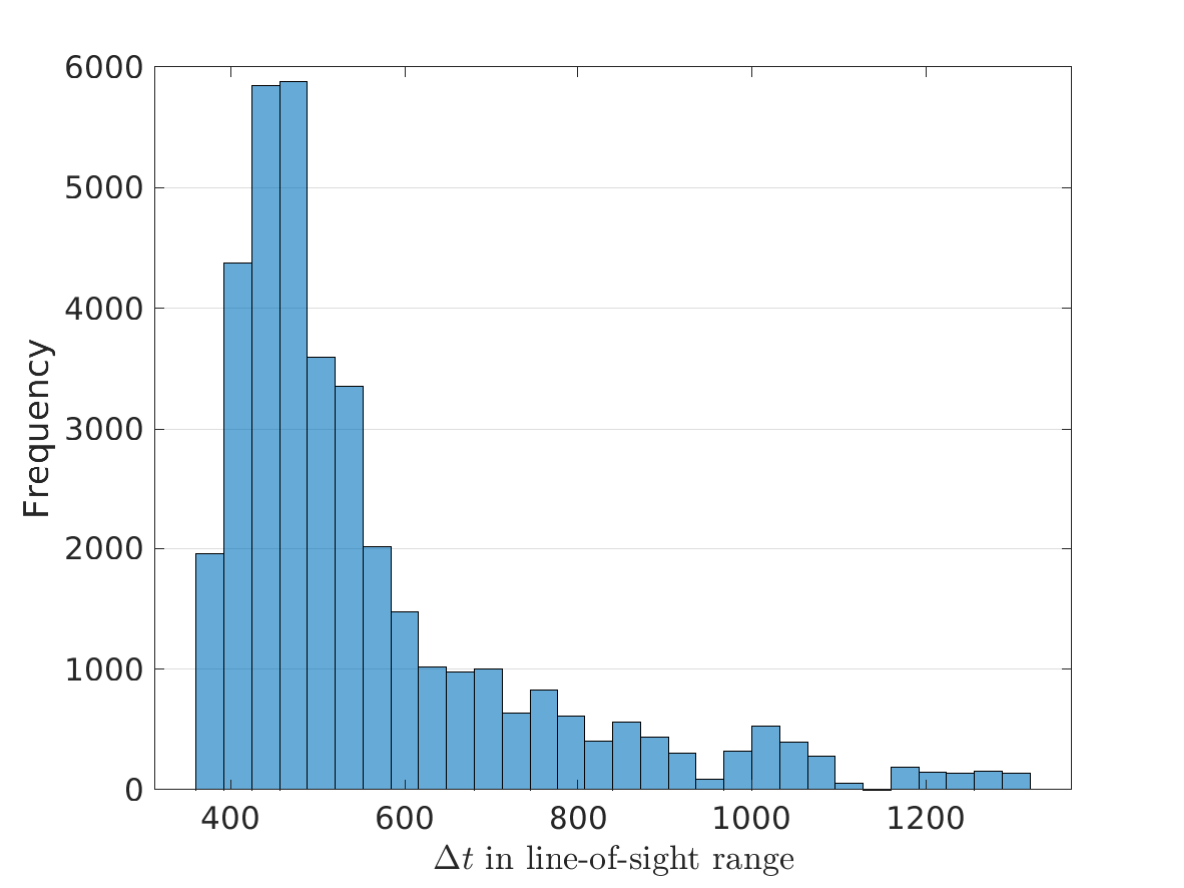}
															\caption{Intervals of time two satellites are in line-of-sight range before establishing a tracking coupling.}
															\label{fig:histogram_links}
														\end{figure}

														The control discretization time is set to $T_c = 10\:\text{s}$, which is small enough to achieve a good approximation of the continuous-time nonlinear dynamics model and large enough such that the control input update frequency is attainable by the Hall effect thrusters. The tracking output coupling topology varies greatly with time, thus the parameters $H$ and $d$ of the scheduling of the RHC distributed and decentralized algorithm have to be tuned thoughtfully, according to the limitations and communication requirements pointed out in Section~\ref{sec:tv_topology}. The interval of time between allowed communications is set to $T_t = 1\:\text{s}$, which is the period of GNSS signals and the sampling time of a filter relying on them. Due to the curvature of the Earth and the low altitude of this LEO shell, the satellites quickly lose line-of-sight between each other. Therefore, at a given time instant, communication between two satellites that are coupled an interval of time later is not necessarily feasible. The approximate theoretical line-of-sight range such that the ISL do not enter the atmosphere any lower than the Thermosphere is given by $R_{LOS} = 2\sqrt{\bar{a}^2-(R_{\bigoplus}+80 \mathrm{km})^2} = 5,014$~km \cite{bhattacherjee2019network}. This range is supported by an ISL system, since multiple laser ISLs of up to 4,900 km have been reported between LEO satellites since 2008 \cite{sodnik2010optical}. To evaluate the volatility of tracking couplings between satellites, a pair of satellites that establishes a tracking coupling in a particular time instant is chosen at random and then the positions of these satellites are backtracked until they are out of line-of-sight range. These procedure is repeated several times. Figure~\ref{fig:histogram_links} depicts an histogram of the interval of time that the pairs of satellites remained in line-of-sight range.

														Consider the computation of the RHC gains over a generic window $\{k,\ldots,k+H-1\}$, according to the scheduling solution proposed in Section~\ref{sec:tv_topology}. The computation of this window starts at $t_s = kT_c - (H+2)T_t$ with the computation of the gains at its end, which corresponds to time instant $t_e = kT_c + HT_c$. If none of the satellites that are coupled at $t_e$ are in line-of-sight at $t_s$, then it is pointless to consider such parameter $H$, since the restricted coupling neighborhood \eqref{eq:D_tv} of each satellite would not contain any other satellites due to communication constraints. Notice that, in Figure~\ref{fig:histogram_links}, no pair of satellites remained in line-of-sight range for more than $\Delta t_{max} = 1320\:\text{s}$. Therefore, the constraint
														\begin{equation}\label{eq:app_constraint_1}
															(H+2)T_t + HT_c < \Delta t_{max}
														\end{equation}
														arises from this observation. Likewise, no restriction due to communication requirements can be enforced for the first $d$ discrete time instants of each RHC finite-window, as concluded in Section~\ref{sec:tv_topology}. The computation of the first $d$ gains of the window starts at  $t_s = kT_c - (d+1)T_t$ with the computation of the gain that corresponds to time instant $t_e = kT_c + dT_c$. To ensure that the communication requirements are met, the satellites that are coupled at $t_e$ must be in line-of-sight, at least, since $t_s$. From Figure~\ref{fig:histogram_links}, the minimum time for the maintenance of line-of-sight is $\Delta t_{min} =360\:\text{s}$. Thus, the constraint
														\begin{equation}\label{eq:app_constraint_2}
															(d+1)T_t + T_cd < \Delta t_{min}.
														\end{equation}
														arises from this observation. After algebraic manipulation of \eqref{eq:app_constraint_1} and \eqref{eq:app_constraint_2} and considering non-overlapping computation windows, as described in Section~\ref{sec:scheduling_ti}, one obtains
														\begin{equation*}
															\begin{cases}
																H < (\Delta t_{max}-2T_t)/(T_t+T_c)\\
																d <  (\Delta t_{min}-T_t)/(T_t+T_c)\\
																d \geq (H+2)T_t/T_c
															\end{cases} \!\!\!\!\!\!\!\!= \begin{cases}
																H < 120.7\\
																d < 32.6 \\
																d \geq 0.2+ H/10 \: .
															\end{cases}
														\end{equation*}
														The parameters $H = 100$ and $d = 25$ were chosen from the rather tight constraints above. Note that these are still valid even if  $T_t$ is doubled.
														
														The weighting matrices $\mathbf{Q}_i^{rel}(k) \in \mathbb{R}^{2\times 2}$, $\mathbf{Q}_i^{in}(k) \in \mathbb{R}^{4\times 4}$, and $\mathbf{R}_i(k) \in \mathbb{R}^{3\times 3}$ were adjusted to physically meaningful orders of magnitude and set to $\mathbf{Q}_i^{rel}(k) = (1/10^{-4})^2 \mathbf{I}$, $\mathbf{Q}_i^{in}(k) = \mathrm{diag}(1/(\bar{a}10^{-4})^2), 1/(0.5\times 10^{-2})^2\mathbf{I}_2, 1/(10^{-2})^2)$, and $\mathbf{R}_i(k) = (1/C_{t1})^2\mathbf{I}_2$.
														
														
														
														\subsection{Simulation results}
														
														In this section, the simulation results are presented for the aforementioned illustrative mega-constellation. {The numerical simulation was carried out employing the} high-fidelity TU Delft's Astrodynamic Toolbox\footnote{TUDAT documentation available at {\small \url{https://docs.tudat.space/}} and source code at {\small \url{https://github.com/tudat-team/tudat-bundle/}}.} (TUDAT) \cite{kumar2012tudat}. {NASA's SPICE ephemerides are used for the orbit propagation and the following perturbations are taken into account:}
														
\begin{figure}[]
															\centering
															\includegraphics[width=\linewidth]{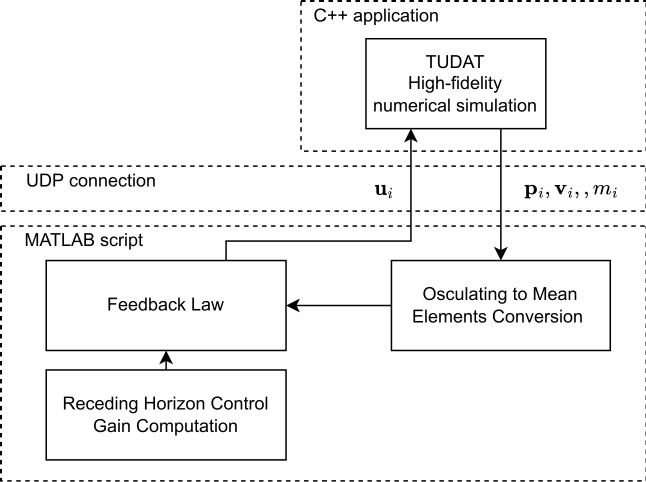}
															\caption{Scheme of the simulation environment.}
															\label{fig:simulation_scheme}
														\end{figure}

														
														\begin{enumerate}
															\item Earth’s gravity field EGM96 spherical harmonic expansion  \cite{lemoine1998development} up to degree and order 24;
															\item Atmospheric drag NRLMSISE-00 model \cite{picone2002nrlmsise}, assuming constant drag coefficient and section area;
															\item Cannon ball solar radiation pressure, assuming constant reflectivity coefficient and radiation area;
															\item Third-body perturbations of the Sun, Moon, Venus, Mars, and Jupiter.
														\end{enumerate} 
														The numerical propagation is assured by a fourth-order Runge-Kutta integration method with fixed step-size of $T_c = 10\:\text{s}$. The feedback control law computation is carried out in MATLAB. The \textit{tudat-matlab-thrust-feedback} package, available at {\small\url{https://github.com/decenter2021/tudat-matlab-thrust-feedback}}, is employed to implement the thrust feedback interface between TUDAT and MATLAB. A scheme of the simulation environment is depicted in Figure~ \ref{fig:simulation_scheme}.

														\begin{figure}[ht!]
															\centering
															\begin{minipage}{0.99\linewidth}
																\centering
																\includegraphics[width=\linewidth]{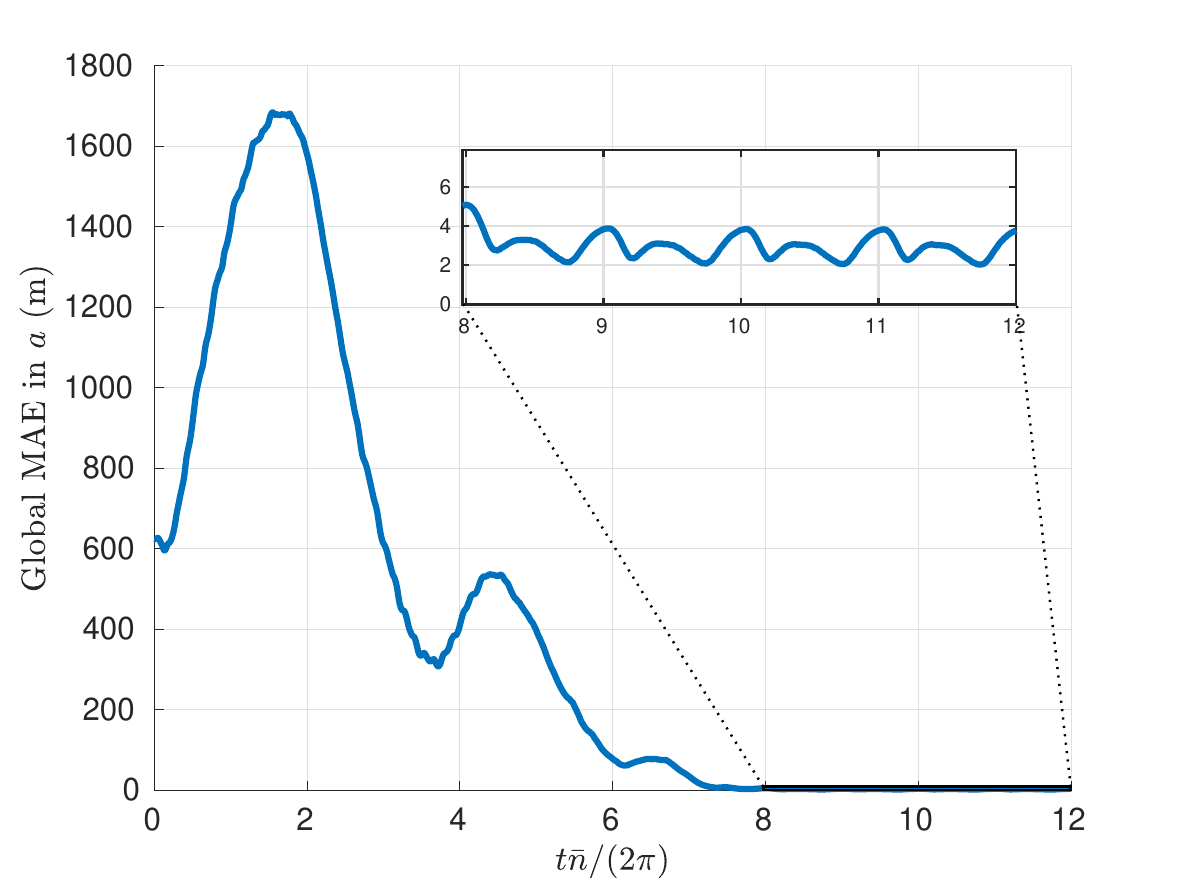}
																\caption{Evolution of the MAE in the semi-major axis.}
																\label{fig:mae_a}
															\end{minipage}\\ \bigskip
															\begin{minipage}{\linewidth}
																\centering
																\includegraphics[width=.99\linewidth]{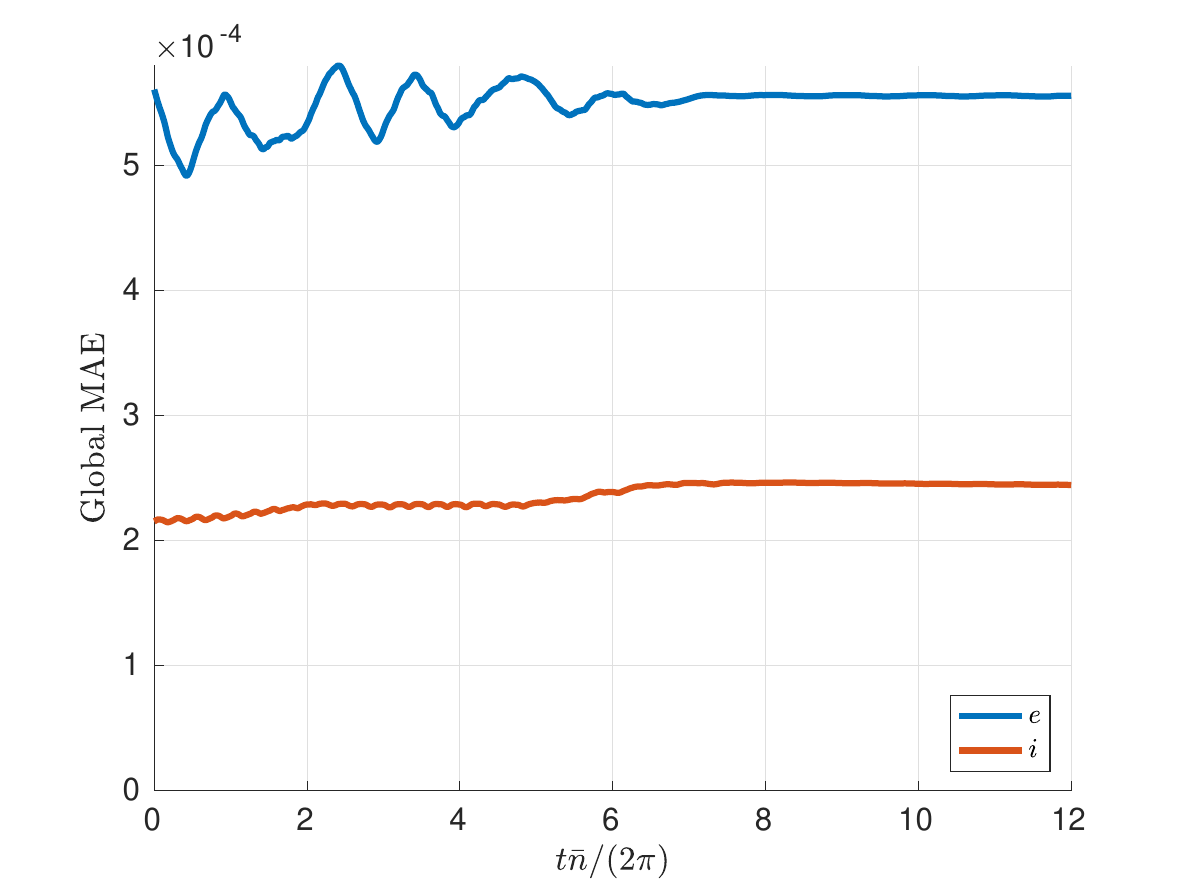}
																\caption{Evolution of the MAE in eccentricity and inclination.}
																\label{fig:mae_ei}
															\end{minipage}\\ \bigskip
															\begin{minipage}{\linewidth}
																\centering
																\includegraphics[width=.99\linewidth]{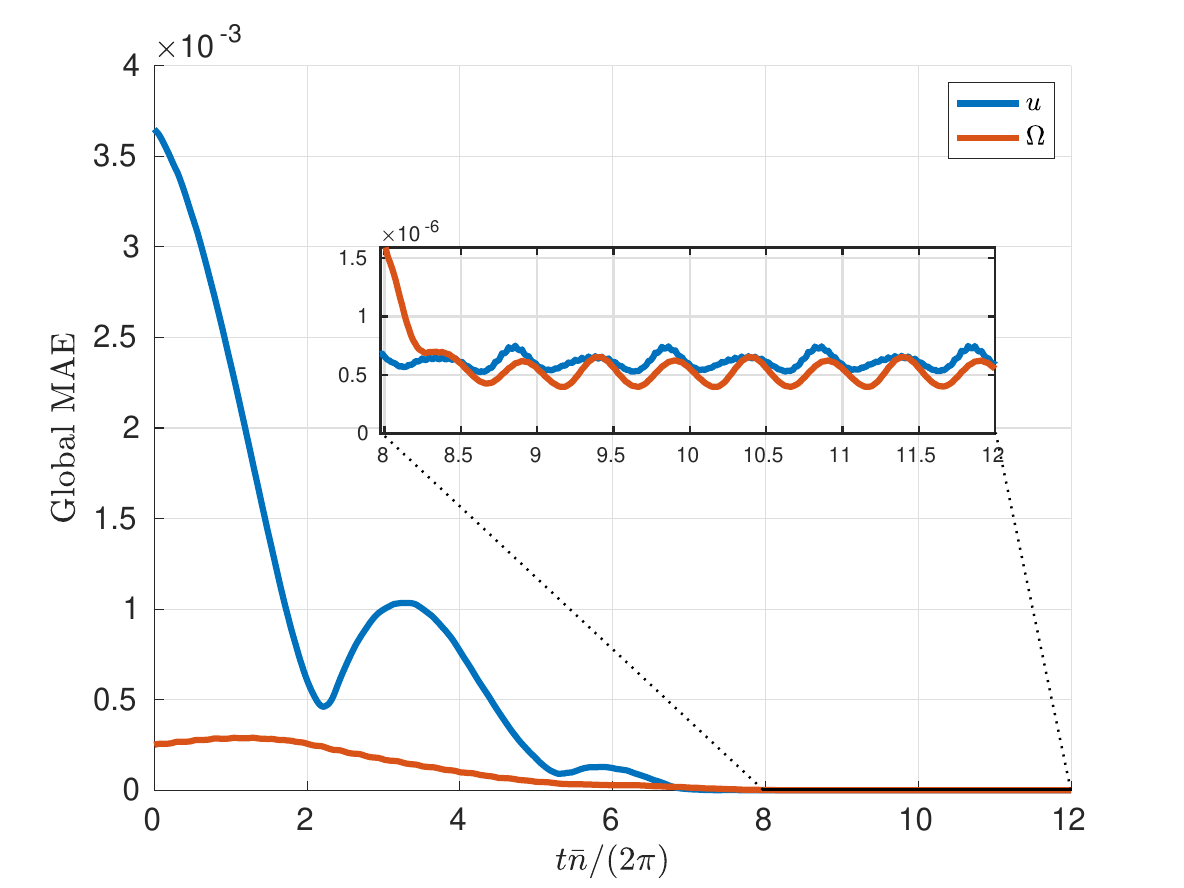}
																\caption{Evolution of the MAE in mean argument of latitude and longitude of ascending node.}
																\label{fig:mae_uOmega}
															\end{minipage}
														\end{figure}

														The total number of states in this shell of the mega-constellation amounts to $6\times 1584 = 9504$, which is the dimension of the global system of an equivalent centralized framework. Hence, implementing a centralized control algorithm in real-time would require the manipulation of very high-dimensional matrices. For example, $\mathbf{P}(\tau)$, which is not typically sparse, would occupy $722.6 \times 10^6$ bytes in double precision. Furthermore, it would have to feature all-to-all communication of substantial data volumes over great distances via the MCC, which requires several ground stations scattered throughout the planet. Therefore, the extreme computational demands in real-time and vast communication requirements make an equivalent centralized framework entirely unfeasible. On top of that, it is also insightful to compare the resource usage of the proposed solution in relation to state-of-the-art decentralized solutions that could, in principle, be used to solve the mega-constellation maintenance problem. First, the decentralized methods that rely on a global synthesis procedure such as  \cite{farhood2015distributed} and \cite{pedroso2021discretecontrol} are required to solve LMIs and perform several algebraic manipulations, respectively, with global matrices of the system dynamics at each discrete-time control instant. However, due to the sheer size of this system, that is unfeasible in a standard computational server. In fact, from the computational analysis of an efficient implementation of the latter method, performed in \cite{PedrosoBatista2021SparseEquation}, one would estimate that each RHC window would take $H\times 3623.7\times 10^{-7}\times9504^{2.770} \approx 3.8\times 10^{9}\, \mathrm{s}$ to be computed. Second, a control design procedure analogous to \cite{luft2018recursive}, whose memory requirements grow with the dimension of the system, would require each individual system to store, just in components of $\mathbf{P}(\tau)$, $1584\times 6^2 \times 8 \approx 45.6 \times 10^4$ bytes in double precision. The proposed solution only requires $|\mathcal{D}^-|_{\text{max}}^2 \times 6^2 \times 8 \approx 1.04 \times 10^4$  bytes in double precision. This reduction of one order of magnitude is especially relevant in stringent settings such satellite design. Moreover, using the design method of \cite{luft2018recursive}, this gap increases significantly as one considers more shells and the memory requirements of each single satellite increase as the number of satellites in the network increases.

														\begin{figure}[ht!]
															\centering
															\includegraphics[width=\linewidth]{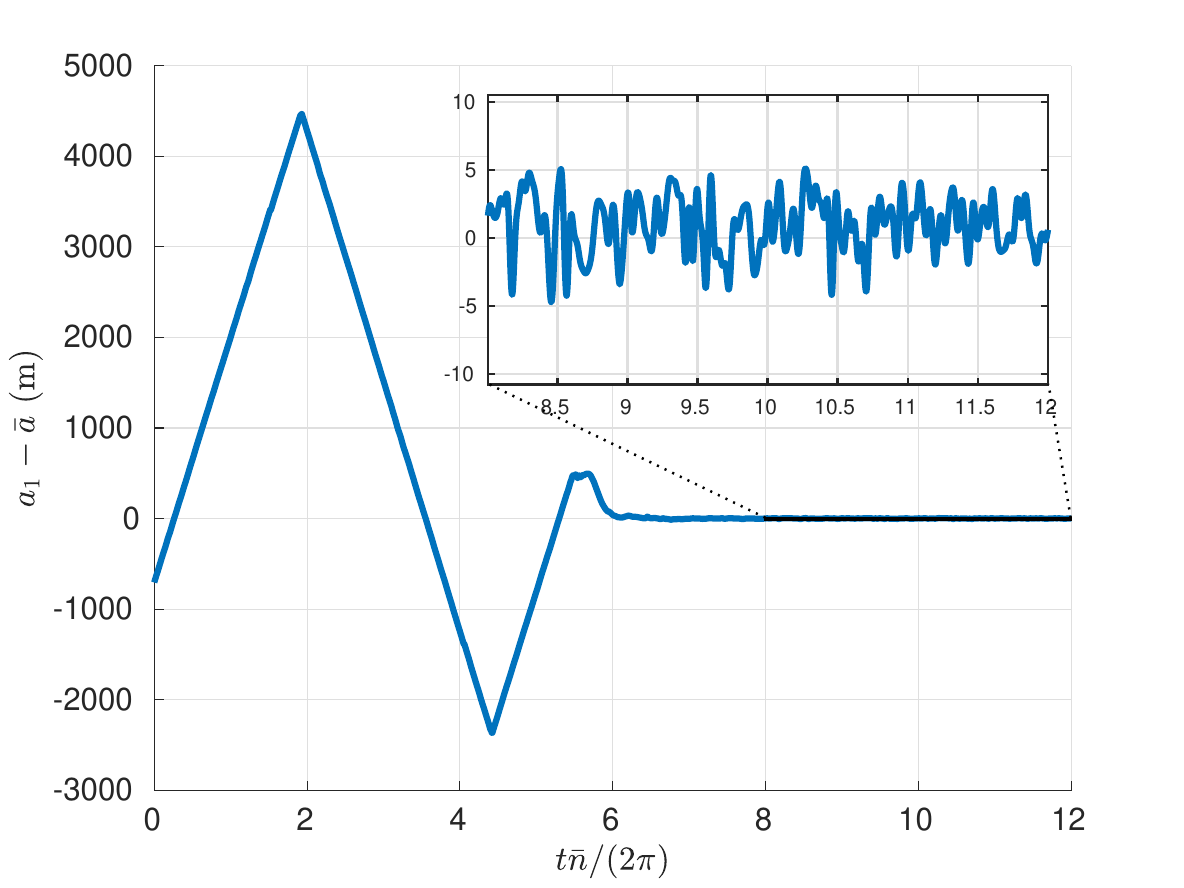}
															\vspace{-0.5cm}
															\caption{Evolution of the semi-major axis tracking error, for satellite 1.}
															\label{fig:asat1}
														\end{figure}
														

														\begin{figure}
															\centering
															\includegraphics[width=\linewidth]{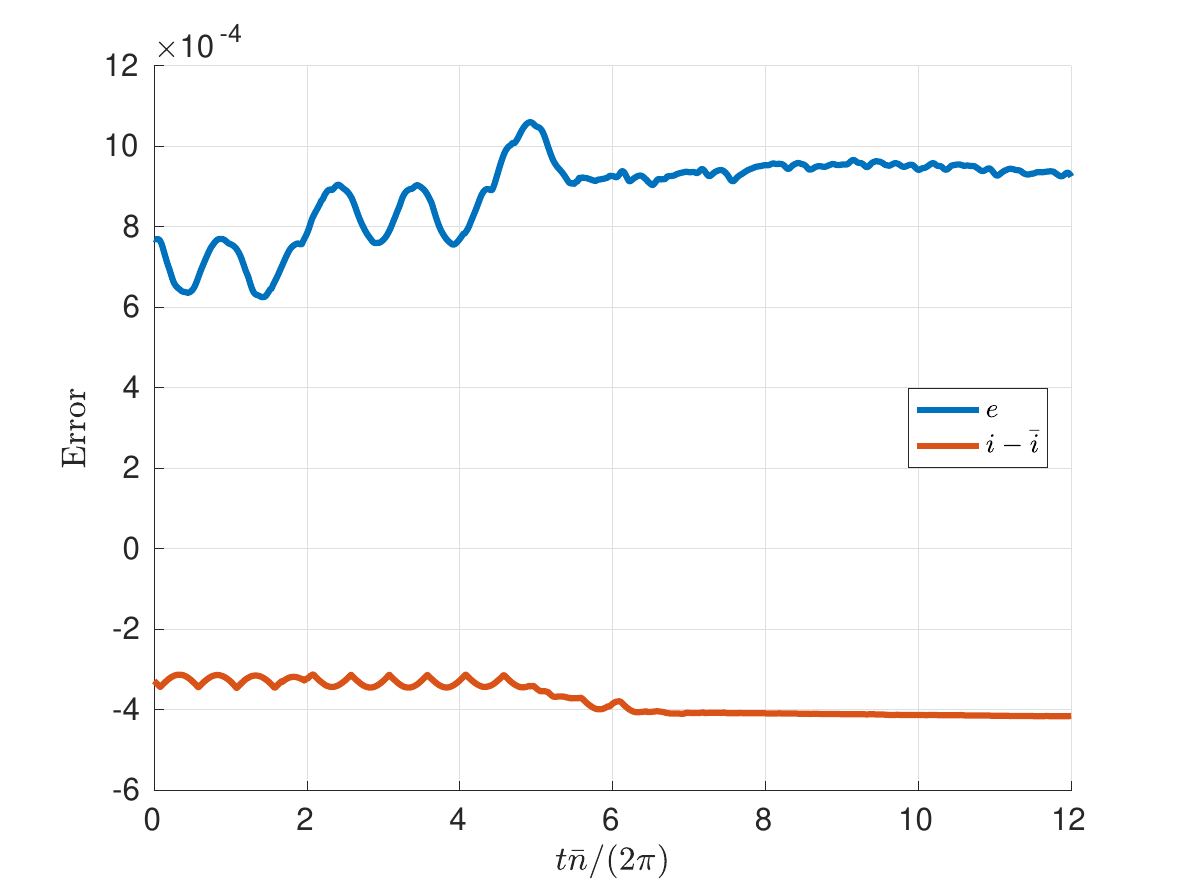}
															\caption{Evolution of the eccentricity and inclination tracking errors, for satellite 1.}
															\label{fig:eisat1}
														\end{figure}
													
														\begin{figure}[t]
															\centering
															\includegraphics[width=\linewidth]{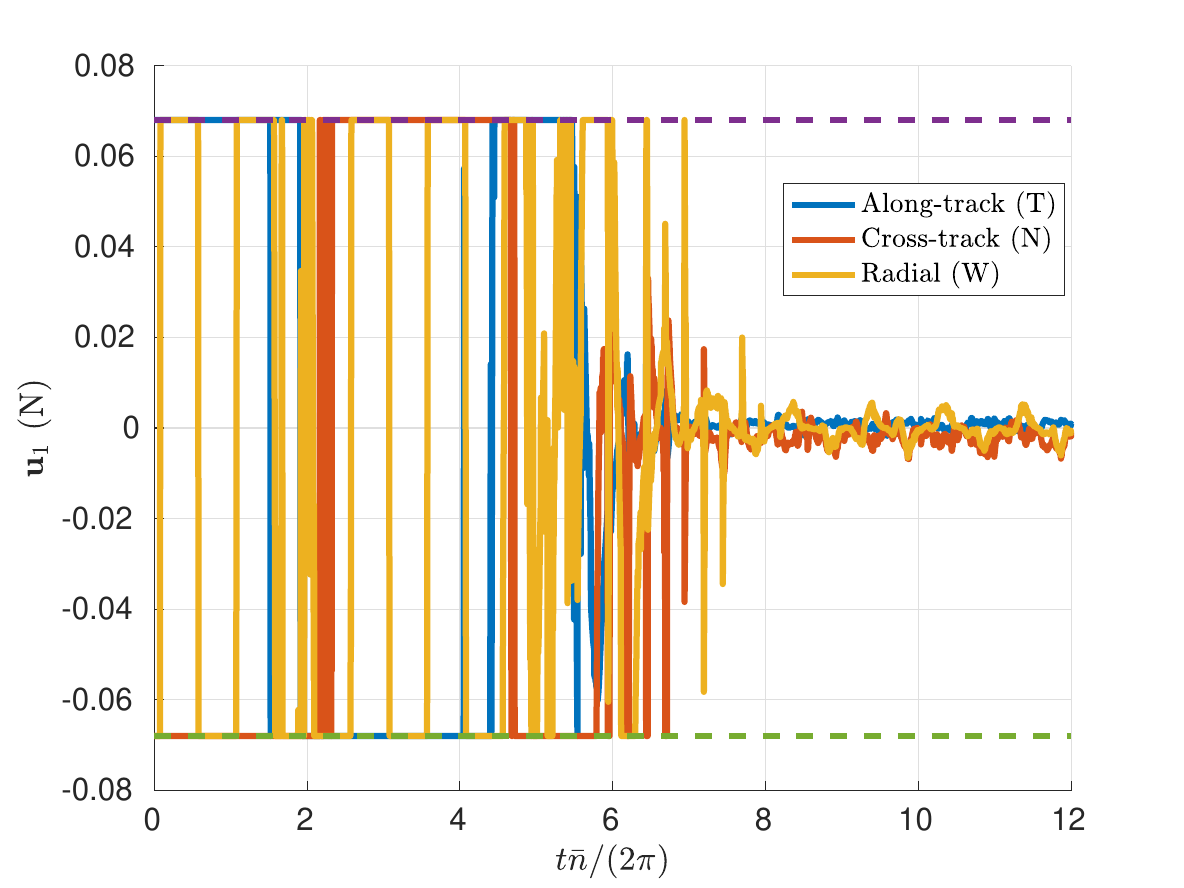}
															\caption{Evolution of the components of the control input, for satellite 1.}
															\label{fig:usat1}
														\end{figure}
													
														\begin{figure}[t]
															\centering
															\includegraphics[width=\linewidth]{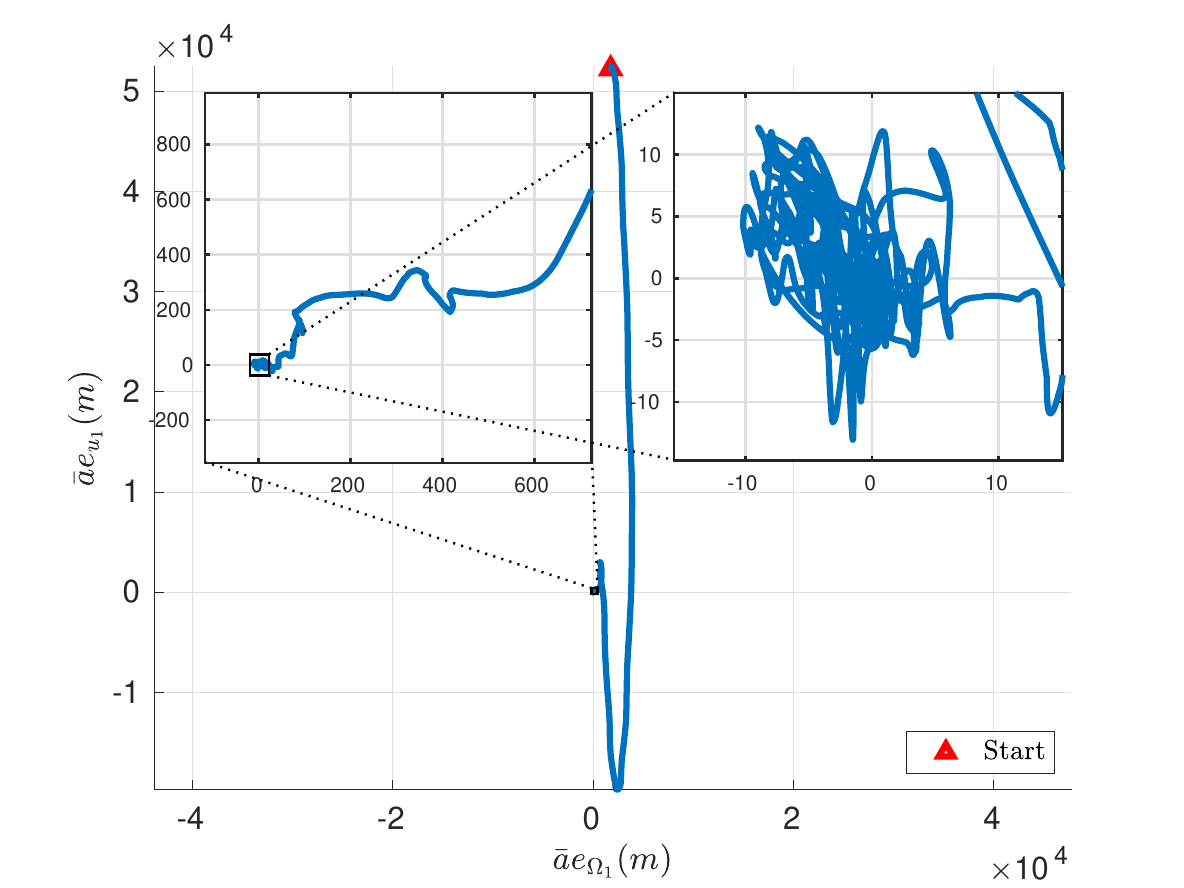}
															\caption{Trajectory of the mean argument of latitude and longitude of ascending node relative tracking errors, for satellite~1.}
															\label{fig:phasedistsat1}
														\end{figure}

														A simulation of the mega-constellation during 12 orbital periods is carried out. An anchor for the nominal constellation is computed at 0 TDB seconds since J2000, according to \eqref{eq:anchor_computation}. The anchor is not updated during the simulation. The evolution of the mean absolute error (MAE) in the semi-major axis, eccentricity, and inclination is depicted in Figures~\ref{fig:mae_a} and \ref{fig:mae_ei}. To evaluate the performance of the relative tracking between the satellites, the mean argument of latitude error, $e_{u_i}(k)$, and longitude of ascending node error, $e_{\Omega_i}(k)$, are defined for each satellite $i$. Consider an instantaneous hypothetical anchor, computed according to \eqref{eq:anchor_computation}, for each time instant $k$. These errors are defined as $e_{u_i}(k) := u_i(k)-\bar{u}_i(k)$ and $e_{\Omega_i}(k) := \Omega_i(k)-\bar{\Omega}_i(k)$, where $\bar{u}_i(k)$ and $\bar{\Omega}_i(k)$ are computed according to \eqref{eq:nominal_orbit} making use of the aforementioned instantaneous hypothetical anchor for time instant $k$. It is very important to remark that these anchors are only employed for performance assessment purposes in a post-processing step, they are not involved in the control law in any way. The evolution of the MAE of the mean argument of latitude and longitude of ascending node is depicted in Figure~\ref{fig:mae_uOmega}. The steady-state MAE, obtained by averaging the MAE of the last three orbital periods of the simulation, is depicted in Table~\ref{tab:ss_mae}.


														First, it is visible that the satellites of the constellation are successfully driven to their nominal semi-major axis and relative separations, despite the large initial errors. Second, although the method proposed in this paper is designed for LTV systems under very strict communication, computational, and memory limitations, it is able to perform well in a network of systems with highly nonlinear dynamics and realistic disturbances. Indeed, the proposed solution is robust to uncertainty, disturbances, and model-reality mismatch. Third, in this simulation there was no need to update the anchor, confirming that its update period is large enough to allow for either a centralized or distributed computation. Fourth, it is visible in Table~\ref{tab:ss_mae} that this solution reaches meter-level accuracy, not only for the semi-major axis, but also for the relative tracking components.

														%
														%
														

														\begin{table*}
															\centering
															\caption{Steady-state MAE.}
															\label{tab:ss_mae}
															\begin{tabular}{lccccc}
																\hline
																& $a-\bar{a}$ (m) & $e$ & $i-\bar{i}$ (rad) & $\bar{a}e_{u_i}$ (m) & $\bar{a}e_{\Omega_i}$ (m)\\ \hline
																Steady-state MAE & 2.887 m & $5.554 \times 10^{-4}$ & $2.450 \times 10^{-4}$ & 4.253 & 3.582 \\ \hline
															\end{tabular}
														\end{table*}

														It is also interesting to analyze the evolution of a single satellite. Figures~\ref{fig:asat1} and \ref{fig:eisat1} show the evolution of the semi-major axis, eccentricity, and inclination tracking errors and Figure~\ref{fig:usat1} depicts the evolution of the components of the control input, all for satellite 1. It is possible to notice that there is a steady-state error in the eccentricity and inclination tracking, but it is not significant. Figure~\ref{fig:phasedistsat1} shows the trajectory of the mean argument of latitude and longitude of ascending node relative tracking errors. It is very interesting to remark that, even for initial kilometer-level relative tracking errors, the proposed solution successfully drives and maintains the shape of the constellation with meter-level accuracy, as seen in Figure~\ref{fig:phasedistsat1}.  Given the very large initial error, the control input is initially saturated, as depicted in Figure~\ref{fig:usat1}. Afterwards, when the relative error is driven to close to zero, the control input is far from the saturation limits. Despite the fact that the proposed method does not have any stability guarantees regarding the saturated inputs, no stability issues arise in the numeric simulation.

														\section{Conclusion}\label{sec:concl}
														Applications of very large-scale networks of dynamically decoupled systems have been emerging in the fields of swarm robotics and networked control, many of which have nonlinear dynamics. The feasible implementation of control algorithms over these networks comes with pressing challenges that arise from the scale of the network, which impose limiting constraints regarding communication, computational power, and memory. State-of-the-art receding horizon control solutions fail to meet such constraints. In this paper, a decentralized receding horizon control solution befitting these challenges is proposed. This algorithm is devised to account for communication and computation delays, only requires local communication and ensures that the computational and memory requirements in each system do not scale with the dimension of the network. As a means of assessing the performance of the algorithm, it is applied to the orbit control problem of low Earth orbit mega-constellations. The proposed method shows promising performance for the orbit control problem of a shell of the Starlink mega-constellation. Future work should focus on devising an approximation to enable a distributed gain synthesis that preserves consistency and on obtaining stability guarantees.

														

														\section*{Code Availability Statement}
														The data that support the findings of this study are openly available in the DECENTER toolbox at {\small \url{https://decenter2021.github.io/examples/DDRHCStarlink}}.


														\section*{Declaration of competing interest}
														The authors declare that they have no known competing financial interests or personal relationships that could have appeared to influence the work reported in this paper.
														
														\section*{Acknowledgments}
														The authors disclosed receipt of the following financial support for the research, authorship, and/or publication of this article: This work was supported by the Funda\c{c}\~ao para a Ci\^encia e a Tecnologia (FCT) through LARSyS - FCT Project UIDB/50009/2020 and through the FCT project DECENTER [LISBOA-01-0145-FEDER-029605], funded by the Programa Operacional Regional de Lisboa 2020 and PIDDAC programs.
														
														\appendix
														\section{Extension to a time-varying tracking output coupling topology}\label{app:alg2}

														The extension of Algorithm~\ref{alg:OSDEKF} to a time-varying tracking output coupling topology is shown in  Algorithm~\ref{alg:OSDEKF_tv}. Consider the in-neighborhood and out-neighborhood $\mathcal{D}_i^-(\tau)$ and $\mathcal{D}_i^+(\tau)$, for $\tau = k,\ldots,k+H$, that would be obtained if no restrictions on the establishment of the communication links existed. Define $\mathcal{C}_i(\tau)$ as the set of systems with which system $\mathcal{S}_i$ can establish a communication link, in an undirected sense, at $t = kT_c - (\tau-k+2)T_t$, for $\tau = k,\ldots,k+H$. Then, consider instead an in-neighborhood and out-neighborhood restricted to the set of systems with which communication is feasible, i.e., 
														\begin{equation}\label{eq:D_tv}
															\tilde{\mathcal{D}}_i^{\pm}\!(\tau) := \mathcal{D}_i^{\pm}(\tau) \cap \mathcal{C}_i(\tau),
														\end{equation}
														for $\tau = k,\ldots,k+H$. Making use of these restricted neighborhoods, the communication requirements in Algorithm~\ref{alg:OSDEKF_tv} are feasible. Note that $\tilde{\phi}(\tau)$ is defined analogously. It is important to point out that to compute the first $d$ gains of the window, which are actually used to compute the control input with \eqref{eq:localFilter}, no restrictions can be enforced, i.e., $\mathcal{D}_i^{\pm}(\tau) \subseteq \mathcal{C}_i(\tau)$. Therefore, $d$ should be sufficiently small to allow for that. In Section~\ref{sec:app_sats}, this scheme is applied to the mega-constellation orbit control problem, for which the communication links between satellites are heavily restricted by the Earth's curvature.
														
														\begin{algorithm*}
															\caption{One-step RHC algorithm for the local gain synthesis of a new window of gains at time instant $k$ in computational unit $\mathcal{T}_i$ for a time-varying coupling topology.}\label{alg:OSDEKF_tv}
															\begin{algorithmic}
																\STATE{\textbf{Output:}} {$\mathbf{K}_{i,p}(\tau), \forall p\in  \tilde{\mathcal{D}}_i^-(\tau), \tau = k,\dots,k+d-1$}
																\STATE \textbf{Step 1}: \textbf{Predict}: $\tilde{\mathcal{D}}_i^-(\tau), \tilde{\mathcal{D}}_i^-(\tau), \tau = k,\ldots,k+H$, according to \eqref{eq:D_tv};\\
																\hspace{2.3cm} $\mathbf{A}_i(\tau), \mathbf{B}_i(\tau)$, $\mathbf{R}_i(\tau), \tau = k,\ldots,k+H-1$;
																\STATE \hspace{2.3cm} $\mathbf{H}_{i,p}(\tau), \forall p \in \tilde{\mathcal{D}}_i^-(\tau), \tau = k+1,\ldots, k+H$;\\
																\hspace{2.3cm} $\mathbf{Q}_{i}(\tau), \tau = k+1,\ldots,k+H$.
																\STATE \textbf{Step 2}: \textbf{Transmit}: $\mathbf{Q}_{i}(k+H)^{1/2}\mathbf{H}_{i,p}(k+H), \forall p\in \tilde{\mathcal{D}}_i^-(k+H)$ to $\forall p\in \tilde{\mathcal{D}}_i^-(k+H) \setminus \{i\}$.\\
																\textbf{Step 3}: \textbf{Receive}: $\mathbf{Q}_{p}(k+H)^{1/2}\mathbf{H}_{p,i}(k+H)$ from $\forall p \in \tilde{\mathcal{D}}^+_i(k+H) \setminus \{i\}$.\\
																\textbf{Step 4}: \textbf{For}: $\tau = k+H-1,\ldots,k$ \\
																\hspace{1.05cm} \textbf{Step 4.1}: \textbf{Transmit}: $\mathbf{R}_i(\tau), \mathbf{B}_i(\tau)$ to $ \forall p \in \tilde{\mathcal{D}}_i^-(\tau)\setminus \{i\}$;\\
																\hspace{4.05cm}$\mathbf{Q}_{p}(\tau+1)^{1/2}\mathbf{H}_{p,i}(\tau+1), \forall p\in \tilde{\mathcal{D}}_i^+(\tau+1)$ to $\forall q\in \tilde{\mathcal{D}}_i^-(\tau) \setminus \{i\}$;\\
																\hspace{4.05cm}\textbf{If:} $\tau \neq k$\\
																\hspace{5.05cm}$\mathbf{Q}_{i}(\tau)^{1/2}\mathbf{H}_{i,p}(\tau), \forall p\in \tilde{\mathcal{D}}_i^-(\tau)$ to $\forall p\in \tilde{\mathcal{D}}_i^-(\tau) \setminus \{i\}$;\\
																\hspace{4.05cm}\textbf{End if}\\
																\hspace{4.05cm}\textbf{If:} $\tau \neq k+H-1$\\
																\hspace{5.05cm}$\mathbf{R}_p(\tau+1), \mathbf{B}_p(\tau+1), \forall p\in \tilde{\mathcal{D}}_i^+(\tau+1)$ to $\forall q\in \tilde{\mathcal{D}}_i^-(\tau) \setminus \{i\}$;\\
																\hspace{5.05cm}$\mathbf{A}_i(\tau+1)$ to $\forall p\in \tilde{\mathcal{D}}_i^-(\tau)$;\\
																\hspace{5.05cm}$\mathbf{K}_{p,i}(\tau+1), \forall p\in \tilde{\mathcal{D}}_i^+(\tau+1)$ to $\forall q\in \tilde{\mathcal{D}}_i^-(\tau) \setminus \{i\}$;\\
																\hspace{5.05cm}$\mathbf{P}_{i,(p,q)}(\tau+1)$ for some $(p,q)\in \phi_i(\tau+1)$ to $\forall l\in \tilde{\mathcal{D}}_i^-(\tau) \setminus \{i\}$.\\	
																\hspace{4.05cm}\textbf{End if}\\
																\hspace{1.05cm} \textbf{Step 4.2}: \textbf{Receive}: $\mathbf{R}_p(\tau), \mathbf{B}_p(\tau)$ from $ \forall p \in \tilde{\mathcal{D}}_i^+(\tau)\setminus \{i\}$;\\
																\hspace{3.85cm}$\mathbf{Q}_{p}(\tau+1)^{1/2}\mathbf{H}_{r,p}(\tau+1), \forall r\in \tilde{\mathcal{D}}_i^+(\tau+1)$ from $\forall p\in \tilde{\mathcal{D}}_i^+(\tau) \setminus \{i\}$;\\
																\hspace{3.85cm}\textbf{If:} $\tau \neq k$\\
																\hspace{4.85cm}$\mathbf{Q}_{p}(\tau)^{1/2}\mathbf{H}_{p,i}(\tau), \forall p\in \tilde{\mathcal{D}}_i^+(\tau)$ from $p\in \tilde{\mathcal{D}}_i^+(\tau) \setminus \{i\}$;\\
																\hspace{3.85cm}\textbf{End if}\\
																\hspace{3.85cm}\textbf{If:} $\tau \neq k+H-1$\\
																\hspace{4.85cm}$\mathbf{R}_r(\tau+1), \mathbf{B}_r(\tau+1), \forall r\in \tilde{\mathcal{D}}_p^+(\tau+1)$ from $\forall p\in \tilde{\mathcal{D}}_i^+(\tau) \setminus \{i\}$;\\
																\hspace{4.85cm}$\mathbf{A}_p(\tau+1)$ from $p\in \tilde{\mathcal{D}}_i^-(\tau)$;\\
																\hspace{4.85cm}$\mathbf{K}_{r,p}(\tau+1), \forall r\in \tilde{\mathcal{D}}_p^+(\tau+1)$ from $\forall p\in \tilde{\mathcal{D}}_i^+(\tau) \setminus \{i\}$;\\
																\hspace{4.85cm}$\mathbf{P}_{p,(r,s)}(\tau+1)$ for some $(r,s)\in \tilde{\phi}_p(\tau+1)$ from $\forall p\in \tilde{\mathcal{D}}_i^+(\tau) \setminus \{i\}$.\\	
																\hspace{3.85cm}\textbf{End if}\\
																\hspace{1.05cm}\textbf{Step 4.3}: \textbf{Compute}:\\
																\hspace{2.45cm}\textbf{If:} $\tau = k +H-1$\\
																\hspace{3.45cm}$\mathbf{P}_{i,(p,q)}(\tau+1)  \gets \!\!\!\!\!\! \sum\limits_{r \in \tilde{\mathcal{D}}^+_p(\tau+1) \cap \tilde{\mathcal{D}}^+_q(\tau+1)}^{\phantom{a}}\mathbf{H}_{r,p}^T(\tau+1)\mathbf{Q}_{r}(\tau+1)\mathbf{H}_{r,q}(\tau+1), \forall (p,q)\in \tilde{\phi}_i(\tau)$;\\
																\hspace{2.45cm}\textbf{Else}\\
																\hspace{3.45cm}$\mathbf{P}_{i,(p,q)}(\tau+1) \gets\!\!\!\!\! \sum\limits_{r \in \tilde{\mathcal{D}}^+_p(\tau+1) \cap  \tilde{\mathcal{D}}^+_q(\tau+1)}^{\phantom{a}}\left(\mathbf{H}_{r,i}^T(\tau+1)\mathbf{Q}_{r}(\tau+1)\mathbf{H}_{r,j}(\tau+1) + \right.$\\ \hspace{3.05cm}$\left.\mathbf{K}_{r,i}^T(\tau+1)\mathbf{R}_{r}(\tau+1)\mathbf{K}_{r,j}(\tau+1)\right)+\tilde{\mathbf{W}}_{p}(k+1)\tilde{\mathbf{P}}_{\tilde{\mathcal{D}}_i^+(\tau+1)}^{(p,q)}(\tau+2)\tilde{\mathbf{W}}_{q}^T(k), \forall (p,q)\in \tilde{\phi}_i(\tau).$\\
																\hspace{2.5cm}\textbf{End if}\\
																\hspace{1.1cm}\textbf{Step 4.4}: \textbf{Compute}:\\
																\hspace{2.4cm} $\mathbf{S}_{p,q}(\tau) \gets \mathbf{B}_p^T(\tau)\mathbf{P}_{i,(p,q)}(\tau+1)\mathbf{B}_{q}(\tau) + \boldsymbol{\delta}_{pq}\mathbf{R}_p(\tau),  \forall (p,q)\in \tilde{\phi}_i(\tau)$; \\
																\hspace{2.4cm} Compute $\tilde{\mathbf{S}}_i(\tau)$ and $\tilde{\mathbf{P}}_i(\tau+1)$ making use of \eqref{eq:def_Stilda} and \eqref{eq:def_Ptilda};\\
																\hspace{2.4cm} $\tilde{\mathbf{K}}_i(\tau) \gets \tilde{\mathbf{S}}_i(\tau)^{-1}\tilde{\mathbf{P}}_i(\tau+1)$;\\
																\hspace{1.1cm}\textbf{End for}\\
																\textbf{Step 5}: \textbf{Transmit}: $\mathbf{K}_{p,i}(\tau), \tau = k,\ldots,k+d-1$ to $\forall p\in {\mathcal{D}}_i^+(\tau)\setminus\{i\}$.\\
																\textbf{Step 6}: \textbf{Receive}: $\mathbf{K}_{i,p}(\tau), \tau = k,\ldots,k+d-1$ from $\forall p\in {\mathcal{D}}_i^-(\tau)\setminus\{i\}$.
															\end{algorithmic}
														\end{algorithm*}

\section*{Funding}
	The authors disclosed receipt of the following financial support for the research, authorship, and/or publication of this article: This work was supported by the Funda\c{c}\~ao para a Ci\^encia e a Tecnologia (FCT) through LARSyS - FCT Project UIDB/50009/2020 and through the FCT project DECENTER [LISBOA-01-0145-FEDER-029605], funded by the Programa Operacional Regional de Lisboa 2020 and PIDDAC programs.

\section*{Supplemental Material}
	The data that support the findings of this study are openly available in the DECENTER toolbox at \href{https://decenter2021.github.io/examples/DDRHCStarlink/}{https://decenter2021.github.io/examples/DDRHCStarlink}.
	
\section*{Declaration of conflicting interests}
	The authors declare that there is no conflict of interest.

\balance
\printbibliography

\end{document}